\documentclass[aps, pre, twocolumn, a4paper, reprint, floatfix, nofootinbib]{revtex4-1}

\usepackage[utf8]{inputenc}
\usepackage{epsfig}
\usepackage[T1]{fontenc}
\usepackage{t1enc}
\usepackage{graphicx}
\usepackage{amssymb}
\usepackage{amsmath}
\usepackage{relsize}
\usepackage{overpic}
\usepackage{bm}
\usepackage{hyperref}

\usepackage[normalem]{ulem}

\newcommand{\avg}[1]{{\left<#1\right>}}

\usepackage{mathtools}
\def\multiset#1#2{\ensuremath{\left(\kern-.3em\left(\genfrac{}{}{0pt}{}{#1}{#2}\right)\kern-.3em\right)}}

\usepackage{amsmath}

\usepackage{verbatim}

\begin{document}

\title{Efficient Monte Carlo and greedy heuristic for the inference of
stochastic block models}

\author{Tiago P. Peixoto}
\email{tiago@itp.uni-bremen.de}
\affiliation{Institut f\"{u}r Theoretische Physik, Universit\"{a}t Bremen, Hochschulring 18, D-28359 Bremen, Germany}


\begin{abstract}
  We present an efficient algorithm for the inference of stochastic
  block models in large networks. The algorithm can be used as an
  optimized Markov chain Monte Carlo (MCMC) method, with a fast mixing
  time and a much reduced susceptibility to getting trapped in
  metastable states, or as a greedy agglomerative heuristic, with an
  almost linear $O(N\ln^2N)$ complexity, where $N$ is the number of
  nodes in the network, independent of the number of blocks being
  inferred. We show that the heuristic is capable of delivering results
  which are indistinguishable from the more exact and numerically
  expensive MCMC method in many artificial and empirical networks,
  despite being much faster. The method is entirely unbiased towards any
  specific mixing pattern, and in particular it does not favor
  assortative community structures.
\end{abstract}

\maketitle

\section{Introduction}

The use of generative models to infer modular structure in networks has
been gaining increased attention in recent
years~\cite{hastings_community_2006, garlaschelli_maximum_2008,
newman_mixture_2007, reichardt_role_2007, hofman_bayesian_2008,
bickel_nonparametric_2009, guimera_missing_2009,
karrer_stochastic_2011,ball_efficient_2011, reichardt_interplay_2011,
zhu_oriented_2012, baskerville_spatial_2011}, due to its more general
character, and because it allows the use of more principled methodology
when compared to more common methods, such as modularity
maximization~\cite{newman_finding_2004}. The most popular generative
model being used for this purpose is the so-called stochastic block
model~\cite{holland_stochastic_1983,fienberg_statistical_1985,
faust_blockmodels:_1992, anderson_building_1992}, where the nodes in the
network are divided into $B$ blocks, and a $B\times B$ matrix specifies
the probabilities of edges existing between nodes of each block. This
simple model generalizes the notion of ``community
structure''~\cite{fortunato_community_2010} in that it accommodates not
only assortative connections, but also arbitrary mixing patterns,
including, for example, bipartite, and core-periphery structures. In
this context, the task of detecting modules in networks is converted
into a process of statistical inference of the parameters of the
generative model given the observed data~\cite{hastings_community_2006,
garlaschelli_maximum_2008, newman_mixture_2007, reichardt_role_2007,
hofman_bayesian_2008, bickel_nonparametric_2009, guimera_missing_2009,
karrer_stochastic_2011,ball_efficient_2011, reichardt_interplay_2011,
zhu_oriented_2012, baskerville_spatial_2011}, which allows one to make
use of the robust framework of statistical analysis. Among the many
advantages which this approach brings is the capacity of separating
noise from structure, such that no spurious communities are
found~\cite{peixoto_parsimonious_2013,
rosvall_information-theoretic_2007,daudin_mixture_2008,
mariadassou_uncovering_2010, moore_active_2011,
latouche_variational_2012, come_model_2013}, increased resolution in
the detection of very small blocks based on refined model selection
methods~\cite{peixoto_hierarchical_2013}, and the identification of
fundamental limits in the detection of modular
structure~\cite{decelle_inference_2011,decelle_asymptotic_2011,
mossel_stochastic_2012, reichardt_detectable_2008,
hu_phase_2012}. However, one existing drawback in the application of
statistical inference is the lack of very efficient algorithms, in
particular for networks with a very large number of blocks, with a
performance comparable to some popular heuristics available for
modularity-based methods~\cite{clauset_finding_2004,
blondel_fast_2008}. Here we present some efficient techniques of
performing statistical inference on large networks, which are partially
inspired by the modularity-based heuristics, but where special care is
taken not to restrict the procedure to purely assortative block
structures, and to control the total number of blocks $B$, such that
detailed model selection criteria can be used. Furthermore, the method
presented functions either as a greedy heuristic, with a fast $O(N\ln^2
N)$ algorithmic complexity, or as full-fledged Monte Carlo method, which
saturates the detectability range of arbitrary modular structure, at the
expense of larger running times.

This paper is divided as follows. In Sec.~\ref{sec:model} the stochastic
block model is defined, together with the maximum likelihood inference
procedure. Sec.~\ref{sec:mcmc} presents an optimized Markov chain Monte
Carlo (MCMC) method which is capable of reaching equilibrium
configurations more efficiently than more unsophisticated approaches. In
Sec.~\ref{sec:heuristic} the MCMC techniques are complemented with an
agglomerative heuristic which successfully avoids metastable states
resulting from starting from random partitions and can be used on its
own as an efficient and high-quality inference method. In this session
we also compare the heuristic to the full MCMC method, for synthetic
networks. In Sec.~\ref{sec:empirical} we compare both methods with
several empirical networks. We finally conclude in
Sec.~\ref{sec:conclusion} with a discussion.

\section{The Stochastic Block Model}\label{sec:model}

The stochastic block model
ensemble~\cite{holland_stochastic_1983,fienberg_statistical_1985,
faust_blockmodels:_1992, anderson_building_1992} is composed of $N$
nodes, divided into $B$ blocks, with $e_{rs}$ edges between nodes of
blocks $r$ and $s$ (or, for convenience of notation, twice that number
if $r=s$). For many empirical networks, much better results are obtained
if degree variability is included inside each block, as in the so-called
degree-corrected block model~\cite{karrer_stochastic_2011}, in which one
additionally specifies the degree sequence $\{k_i\}$ of the graph as an
additional set of parameters.

The detection of modules consists in inferring the most likely model
parameters which generated the observed network. One does this by
finding the best partition $\{b_i\}$ of the nodes, where $b_i \in [1,
B]$ is the block membership of node $i$, in the observed network $G$,
which maximizes the posterior likelihood
$\mathcal{P}(G|\{b_i\})$. Because each graph with the same edge counts
$e_{rs}$ are equally likely, the posterior likelihood is
$\mathcal{P}(G|\{b_i\}) = 1/\Omega(\{e_{rs}\},\{n_r\})$, where $e_{rs}$
and $n_r$ are the edge and node counts associated with the block
partition $\{b_i\}$, and $\Omega(\{e_{rs}\},\{n_r\})$ is the number of
different network realizations. Hence, maximizing the likelihood is
identical to minimizing the microcanonical
entropy~\cite{bianconi_entropy_2009} $\mathcal{S}(\{e_{rs}\},\{n_r\}) =
\ln\Omega(\{e_{rs}\},\{n_r\})$, which can be
computed~\cite{peixoto_entropy_2012} as
\begin{equation}\label{eq:st}
  \mathcal{S}_t = \frac{1}{2} \sum_{rs}n_rn_sH_{\text{b}}\left(\frac{e_{rs}}{n_rn_s}\right),
\end{equation}
for the traditional model and
\begin{equation}\label{eq:sc}
  \mathcal{S}_c \simeq -E -\sum_kN_k\ln k! - \frac{1}{2} \sum_{rs}e_{rs}\ln\left(\frac{e_{rs}}{e_re_s}\right),
\end{equation}
for the degree corrected variant, where $E=\sum_{rs}e_{rs}/2$ is the
total number of edges, $N_k$ is the total number of nodes with degree
$k$, $e_r=\sum_se_{rs}$ is the number of half-edges incident on block
$r$, and $H_{\text{b}}(x) = -x\ln x - (1-x)\ln (1-x)$ is the binary
entropy function, and it was assumed that $n_r \gg 1$.

These models can be generalized for directed networks, for which
corresponding expressions for the entropies are easily
obtained~\cite{peixoto_entropy_2012, peixoto_parsimonious_2013}. The
methods described in this paper are directly applicable for directed
networks as well.

Although minimizing $\mathcal{S}_{t/c}$ allows one to find the most
likely partition into $B$ blocks, it cannot be used to find the best
value of $B$ itself. This is because the minimum of $\mathcal{S}_{t/c}$
is a strictly decreasing function of $B$, since larger models can always
incorporate more details of the observed data, providing a better
fit. Indeed, if one minimizes $\mathcal{S}_{t/c}$ over all $B$ values
one will always obtain the trivial $B=N$ partition where each node is in
its own block, which is not a useful result. The task of identifying
the best value of $B$ in a principled fashion is known as model
selection, which attempts to separate actual structure from noise and
avoid overfitting. In the current context this can be done in a variety
of ways, such as using the minimum description length (MDL)
criterion~\cite{rosvall_information-theoretic_2007,
peixoto_parsimonious_2013} or performing Bayesian model selection (BMS)
~\cite{daudin_mixture_2008, guimera_missing_2009,
mariadassou_uncovering_2010, moore_active_2011,
latouche_variational_2012, come_model_2013}. In
Ref.~\cite{peixoto_hierarchical_2013} a high-resolution model selection
method is presented, which is based on MDL and a hierarchy of nested
stochastic block models describing the network topology at multiple
scales and is capable of discriminating blocks with sizes significantly
below the so-called ``resolution limit'' present in other model
selection procedures, and other community detection heuristics such as
modularity optimization~\cite{fortunato_resolution_2007}. In
Ref.~\cite{peixoto_hierarchical_2013} it is also shown that BMS and MDL
deliver identical results if the same model constraints are
imposed. However, in order to perform model selection, one first needs
to find optimal partitions of the network for given values of $B$, which
is the subproblem which we consider in detail in this work. Therefore,
in the remainder of this paper we will assume that the value of $B$ is a
fixed parameter, unless otherwise stated, but the reader should be aware
that this value itself can be determined at a later step via model
selection, as described e.g. in Refs.~\cite{peixoto_parsimonious_2013,
peixoto_hierarchical_2013}.

Given a value of $B$, directly obtaining the partition $\{b_i\}$ which
minimizes $\mathcal{S}_{t/c}$ is in general not tractable, since it
requires testing all possible partitions, which is only feasible for
very small networks. Instead one must rely on approximate, or stochastic
procedures which are guaranteed to sample partitions with a probability
given as a function of $\mathcal{S}_{t/c}$, as described in the
following section.

\section{Markov Chain Monte Carlo}\label{sec:mcmc}

\begin{figure}

  \includegraphics{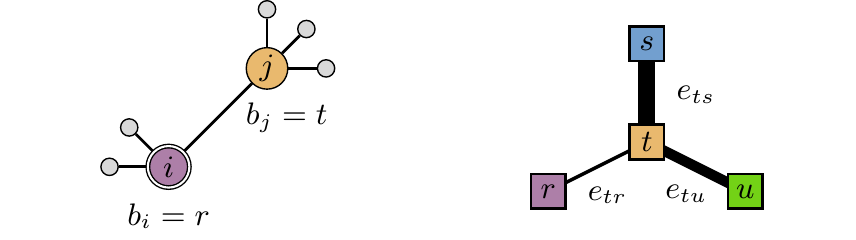}

  \caption{\label{fig:move}\emph{Left:} Local neighborhood of node $i$
  belonging to block $r$, and a randomly chosen neighbor $j$ belonging
  to block $t$. \emph{Right:} Block multigraph, indicating the number of
  edges between blocks, represented as the edge thickness. In this
  example, the attempted move $b_i \to s$ is made with a larger
  probability than either $b_i \to u$ or $b_i \to r$ (no movement),
  since $e_{ts} > e_{tu}$ and $e_{ts} > e_{tr}$.}
\end{figure}

The MCMC approach consists in modifying the block membership of each
node in a random fashion and accepting or rejecting each move with a
probability given as a function of the entropy difference $\Delta
S_{t/c}$. If the acceptance probabilities are chosen appropriately and
the process is ergodic, i.e., all possible network partitions are
accessible, and detailed balance is preserved,
i.e., the moves are reversible, after a sufficiently long equilibration
time, each observed partition must occur with the desired probability
proportional to $\mathcal{P}(G|\{b_i\}) = e^{-S_{t/c}}$. In this
sense, this process is exact, since it is guaranteed to eventually
produce the partitions with the desired probabilities, after a
sufficient long equilibration (or mixing) time. In practice, the
situation is more nuanced, since equilibration times may be very long,
and one may not able to sample from a good approximation of the desired
distribution, and different ways of implementing the Markov chain leads
to different mixing times. The simplest approach one can take is to
attempt to move each vertex into one of the $B$ blocks with equal
probability. This easily satisfies the requirements of ergodicity and
detailed balance, but can be very inefficient. This is particularly so
in the case where the value of $B$ is large, and the block structure of
the network is well defined, such that the vertex will belong to very
few of the $B$ blocks with a non-vanishing probability, which means that
most random moves will simply be rejected. A better approach has been
proposed in Ref.~\cite{peixoto_parsimonious_2013}, which we present here in a
slightly generalized fashion, and consists in attempting to move a
vertex from block $r$ to $s$ with a probability given by
\begin{equation}\label{eq:move}
  p(r\to s|t) = \frac{e_{ts} + \epsilon}{e_t + \epsilon B},
\end{equation}
where $t$ is the block label of a randomly chosen neighbor, and
$\epsilon > 0$ is a free parameter (note that by making $\epsilon \to
\infty$ we recover the fully random moves described
above). Eq.~\ref{eq:move} means that we attempt to guess the block
membership of a given node by inspecting the block membership of its
neighbors and by using the currently inferred model parameters to choose
the most likely blocks to which the original node belongs (see
Fig.~\ref{fig:move}). It should be observed that this move imposes no
inherent bias; in particular, it does not attempt to find assortative
structures in preference to any other, since it depends fully on the
matrix $e_{rs}$ currently inferred. For any choice of $\epsilon > 0$,
this move proposal fulfills the ergodicity condition, but not detailed
balance. However, this can be enforced in the usual Metropolis-Hastings
fashion~\cite{metropolis_equation_1953,hastings_monte_1970} by accepting
each move with a probability $a$ given by
\begin{equation}\label{eq:a}
a = \min\left\{e^{-\beta\Delta \mathcal{S}_{t/c}} \frac{\sum_tp_t^ip(s\to r|t)}{\sum_tp^i_tp(r \to s|t)}, 1\right\},
\end{equation}
where $p^i_t$ is the fraction of neighbors of node $i$ which belong to
block $t$, and $p(s\to r|t)$ is computed after the proposed $r\to s$
move (i.e., with the new values of $e_{rt}$), whereas $p(r\to s|t)$ is
computed before. The parameter $\beta$ in Eq.~\ref{eq:a} is an inverse
temperature, which can be used to escape local minima or to turn the
algorithm into a greedy heuristic, as discussed below.

The moves with probabilities given by Eq.~\ref{eq:move} can be
implemented efficiently. We simply write $p(r\to s|t) =
(1-R_t)e_{ts}/e_t + R_t/B$, with $R_t = \epsilon B / (e_t + \epsilon
B)$. Hence, in order to sample $s$ we proceed as follows:
1. A random neighbor $j$ of the node $i$ being moved is selected, and
its block membership $t=b_j$ is obtained; 2. The value $s$ is randomly
selected from all $B$ choices with equal probability; 3. With
probability $R_t$ it is accepted; 4. If it is rejected, a randomly
chosen edge adjacent to block $t$ is selected, and the block label $s$
is taken from its opposite endpoint. This simple procedure selects the
value of $s$ with a probability given by $\sum_tp^i_tp(r \to s|t)$, and
requires only a small number of operations, which is independent either
on $B$ or the number of neighbors the node $i$ has. The only requirement
is that we keep a list of edges which are adjacent to each block, which
incurs an additional memory complexity of $O(E)$. To decide whether to
accept the move, we need to compute the value of $a$, which can be done
in $O(k_i)$ time, which is the same number of operations which is
required to compute $\Delta S_{t/c}$~\footnote{For sparse networks with
$e_{rs} \ll n_rn_s$, we may write $\mathcal{S}_t \cong E - \frac{1}{2}
\sum_{rs}e_{rs}\ln e_{rs} + \sum_re_r\ln{n_r}$, and note that to
compute the change in entropy we need to modify at most $4k$ terms in
the first sum and $2$ terms in the second, if we change the membership
of a node with degree $k$. The same argument holds for
$\mathcal{S}_c$.}. Therefore, an entire MCMC sweep of all nodes in the
network requires $O(E)$ operations, independent of $B$.

\begin{figure}
  \includegraphics[width=.49\columnwidth]{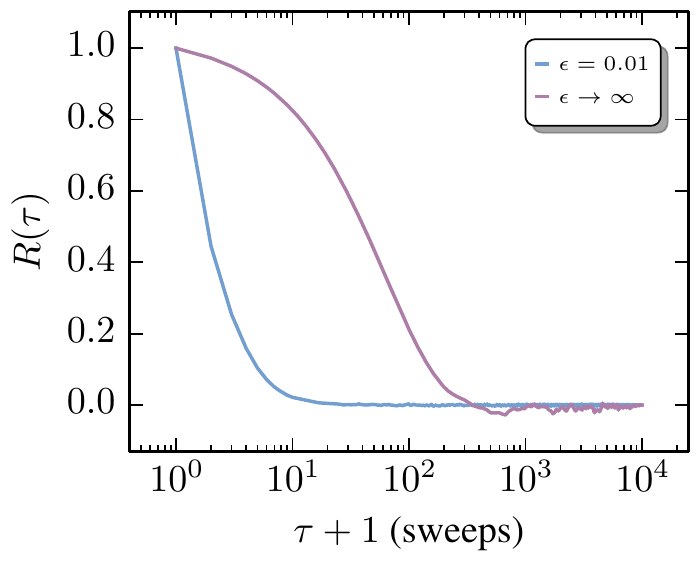}
  \includegraphics[width=.49\columnwidth]{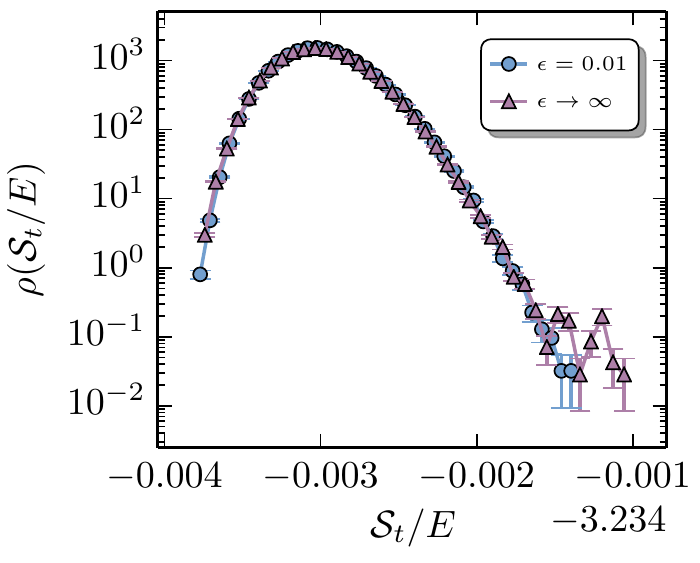}

  \caption{\label{fig:mix}
    Left: Autocorrelation function $R(\tau)$, for a PP model with $c=0.8$
    and $B=100$, for a network of size $N=10^4$ and $\avg{k} = 10$, and
    two values of the parameter $\epsilon$, where for
    $\epsilon\to\infty$ we have fully random moves. The curves were
    averaged for $100$ independent network realizations. Right: PDF of
    the values of $\mathcal{S}_t / E$ obtained for $T=2\times 10^4$
    consecutive sweeps for $100$ independent network realizations, for
    different $\epsilon$ values, showing the same distribution.}
\end{figure}

To test the behavior of this approach, we examine a simple example known
as the Planted Partition (PP) model~\cite{condon_algorithms_2001}. It
corresponds to an assortative block structure given by
$e_{rs}=2E[\delta_{rs}c / B + (1-\delta_{rs})(1-c) / B(B-1)]$,
$n_r=N/B$, and $c\in[0, 1]$ is a free parameter which controls the
assortativity strength. In this example, the algorithm above leads to
much faster mixing times, as can be seen in Fig.~\ref{fig:mix}(left), which
shows the autocorrelation function
\begin{equation}
R(\tau)=\frac{\sum_{t=1}^{T-\tau}
\left(\mathcal{S}_{t/c}(t)-\avg{\mathcal{S}_{t/c}}\right)\left(\mathcal{S}_{t/c}(t+\tau)-\avg{\mathcal{S}_{t/c}}\right)}{{(T-\tau)\sigma_{\mathcal{S}_{t/c}}^2} },
\end{equation}
where $\mathcal{S}_{t/c}(t)$ is the entropy value after $t$ MCMC sweeps,
and $T$ is the total number of sweeps, computed after a sufficiently
long transient has been discarded. For the particular choice of
parameters chosen for Fig.~\ref{fig:mix}, the autocorrelation time is of
the order of $10$ sweeps with the optimized moves, and of the order of
$100$ sweeps with the fully random variant. Despite the difference in
the mixing time, both methods sample from the same distribution, as
shown in Fig.~\ref{fig:mix}(right).

\begin{figure}
  \includegraphics[width=.49\columnwidth]{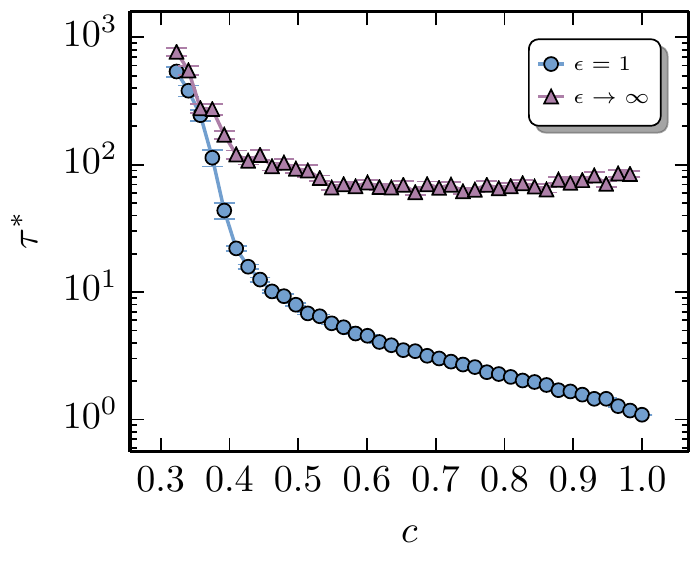}
  \includegraphics[width=.49\columnwidth]{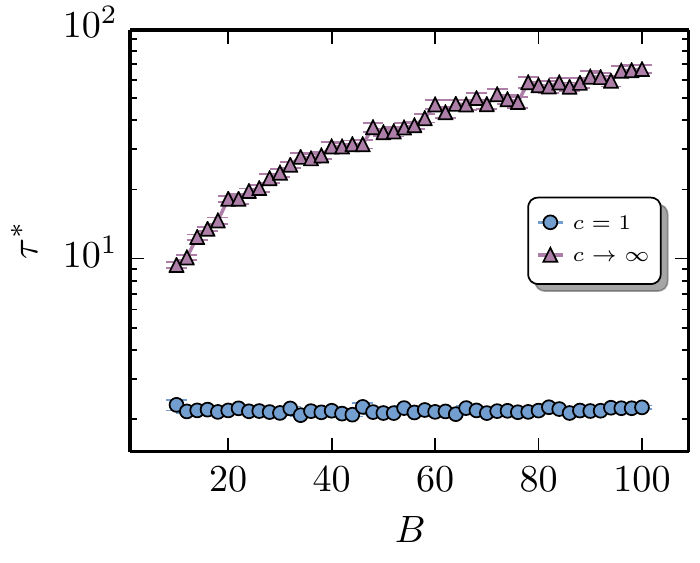}

  \caption{\label{fig:mix_cB} Left: Correlation time $\tau^*$ as a
  function of the model parameter $c$, for different values of
  $\epsilon$, $N=10^4$, $\avg{k}=10$, $B=100$, averaged over $40$
  independent network realizations. Right: Correlation time $\tau^*$ as
  a function of the number of blocks $B$, for different values of
  $\epsilon$, for $N=100 \times B$, $\avg{k}=10$, $c=0.8$, averaged
  over $40$ independent network realizations.}
\end{figure}

The improvement for smaller $\epsilon$ values is more prominent as the
block structure becomes more well-defined, as can be seen in
Fig.~\ref{fig:mix_cB}, which shows the autocorrelation time $\tau^*$,
defined here as
\begin{equation}
  \tau^*=\sum_{\tau=0}^{T'}R(\tau),
\end{equation}
where $T'$ is the largest value of $\tau$ for which $R(\tau) \ge 0$.  In
Fig.~\ref{fig:mix_cB}(left) are shown the values of $\tau^*$ depending
on $c$, from which one can see that the relative improvement on the
mixing time can be up to two orders of magnitude, for the chosen value
of $B=100$. As the value of $c$ approaches the detectability threshold
(see below), the autocorrelation time diverges, as is typical of
second-order phase transitions, and the relative advantage of the
optimized moves diminishes. However, for most of the parameter range
where the blocks are detectable, the mixing time with the optimized
moves seems independent on the actual number of blocks, as shown in
Fig.~\ref{fig:mix_cB}, where a fixed block size $N/B = 100$ was used,
and $B$ was varied. One can see that for the optimized moves the mixing
time remains constant, whereas for the fully random moves it increases
steadily with $B$.

\begin{figure}
  \includegraphics[width=1\columnwidth]{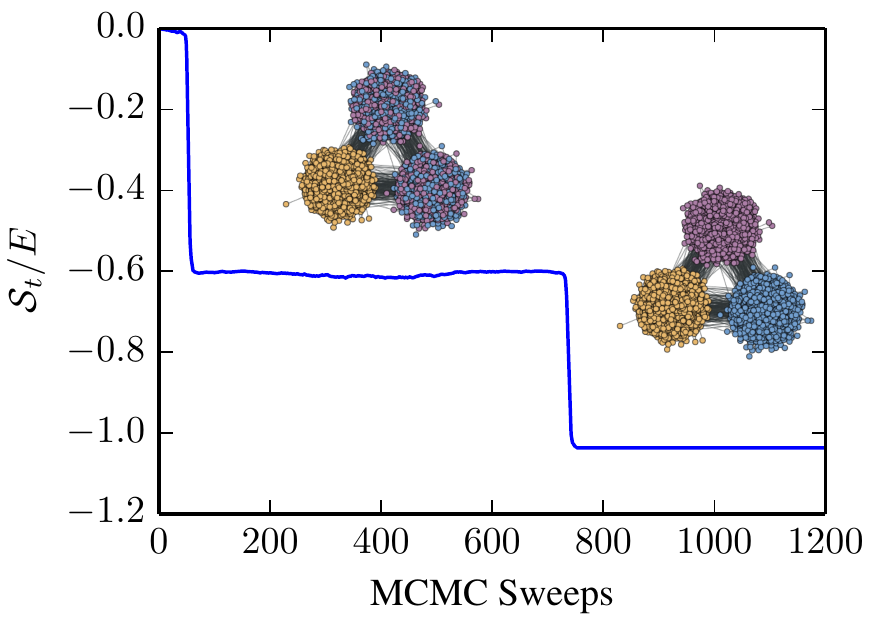}

  \caption{\label{fig:meta} Evolution of the MCMC for a network sampled
  from the PP model with $N=10^4$, $\avg{k}=10$, $B=3$ and $c=0.99$,
  starting from a fully random partition of the nodes. The networks show
  a representative snapshot of the state of the system before and after
  the last drop in $\mathcal{S}_t$.
 }
\end{figure}

Although the optimized moves above provide a considerable improvement
over the fully random alternative whenever the number of blocks $B$
becomes large, there remains an important problem when applying
it. Namely, the mixing time can be heavily dependent on how close one
starts from the typical partitions which are obtained after
equilibration. Since one does not know this, one often starts with a
random partition. However, this is very far from the equilibrium states,
and if the block structure is sufficiently strong, this can lead to
metastable configurations, where the block structure is only partially
discovered, as shown in Fig.~\ref{fig:meta}, for a network with
$B=3$~\footnote{The occurrence of these metastable states is independent
of the optimized moves and happens also for the fully random
$\epsilon\to\infty$ moves.}. The main problem is that not only does it
take a long time to escape such metastable states, but also by observing
the values of $S_{t/c}$ alone, one may arrive at the wrong conclusion
that the Markov chain has equilibrated. For example, in the simulation
shown in Fig.~\ref{fig:meta}, it took many hundreds of sweeps for the
final drop in $\mathcal{S}_t$ to occur, and before this, the time series
is difficult to distinguish from an equilibrated chain.  This problem is
exacerbated if the average block size $N/B$ increases, which can be
frustrating since one would like to consider these scenarios to be
easier than for smaller block sizes. In order to avoid this problem, we
propose the agglomerative heuristic described in the next session, which
can be used as a privileged starting point for the Markov chain, or as
an approximate inference tool on its own.

\section{Agglomerative Heuristic}\label{sec:heuristic}

\begin{figure}
  \includegraphics{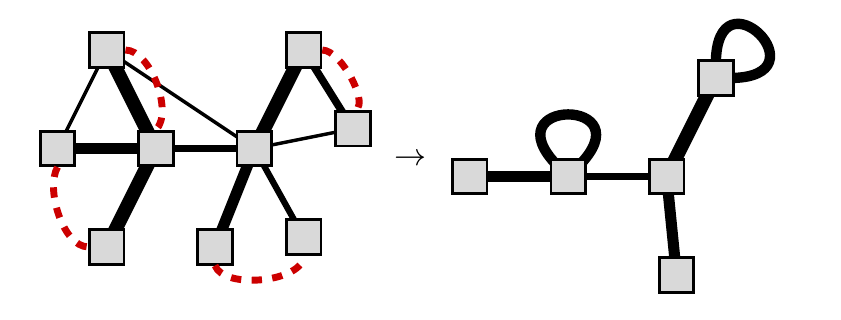}

  \caption{\label{fig:merge} Representation of the block merges used in
  the agglomerative heuristic. Each square node is a block in the original
  graph, and the merges (represented as red dashed lines) correspond
  simply to block membership moves.}
\end{figure}

In order to avoid the metastable states described previously, we explore
the fact that they are more likely to occur if the block sizes are
large, since otherwise the quenched topological fluctuations present in
the network will offer a smaller free-energy barrier which needs to be
overcome. Therefore, a more promising approach is to attempt to find the
best configuration for some $B' > B$, and then use that configuration to
obtain a better estimate for one with $B$ blocks~\footnote{Note that we
cannot simply set $B'>B$ and perform the same MCMC sweeps, expecting to
obtain a partition into $B$ blocks, since the values of
$\mathcal{S}_{t/c}$ obtained for larger $B$ values are always
smaller. Differently from other community detection approaches such as
modularity optimization, here we are forced to control the value of $B$
explicitly, which we can determine at a later step via a model selection
procedure, as discussed previously.}. This can be done by merging blocks
together progressively, as shown in Fig.~\ref{fig:merge}. We implement
this by constructing a block (multi)graph, where the blocks themselves
are the nodes (weighted by the block sizes) and the edge counts $e_{rs}$
are the edge multiplicities between each block node. In this
representation, a block merge is simply a block membership move of a
block node, where initially each node is in its own block. The choice of
moves is done with same probability as before,
i.e. via Eq.~\ref{eq:move}. In order to select the best merges, we
attempt $n_m$ moves for each block node, and collectively rank the best
moves for all nodes according to $\Delta S_{t/c}$. From this global
ranking, we select the best $B' - B$ merges to obtain the desired
partition into $B$ blocks. However if the value of $N/B'$ itself is too
large, we face again the same problem as before. Therefore we proceed
iteratively by starting with $B_1 = N$, and selecting $B_{i+1} = B_i /
\sigma$, until we reach the desired $B$ value, where $\sigma > 1$
controls how greedily the merges are performed.  To diminish the effect
of bad merges done in the earlier steps, we also allow individual node
moves between each merge step, by applying the MCMC steps above to the
original network, with $\beta\to\infty$.  The complexity of each
agglomerative step is $O[n_mE
+ N\ln(B_i - B_{i - 1})
+ \tau E]$, which incorporates the search for the merge candidates, the
ranking of the $B_i - B_{i - 1}$ best merges, and the movement of the
individual nodes, where $\tau$ is the necessary amount of sweeps to
reach a local minimum. Since we have in total $\ln (N / B) / \ln \sigma$
merge steps, with the slowest one being the first with $B_1=N$, we have
an overall complexity of $O\{[(n_m + \tau) E + N\ln N] \times \ln N /
\ln\sigma\} \sim O(N\ln^2N)$, if we assume that $B \ll N$~\footnote{This
is a worst-case scenario. If $B \sim N$, then the complexity reduces to
$O(N\ln N)$.} and that the graph is sparse with $E \sim O(N)$.

\begin{figure}
  \includegraphics[width=.49\columnwidth]{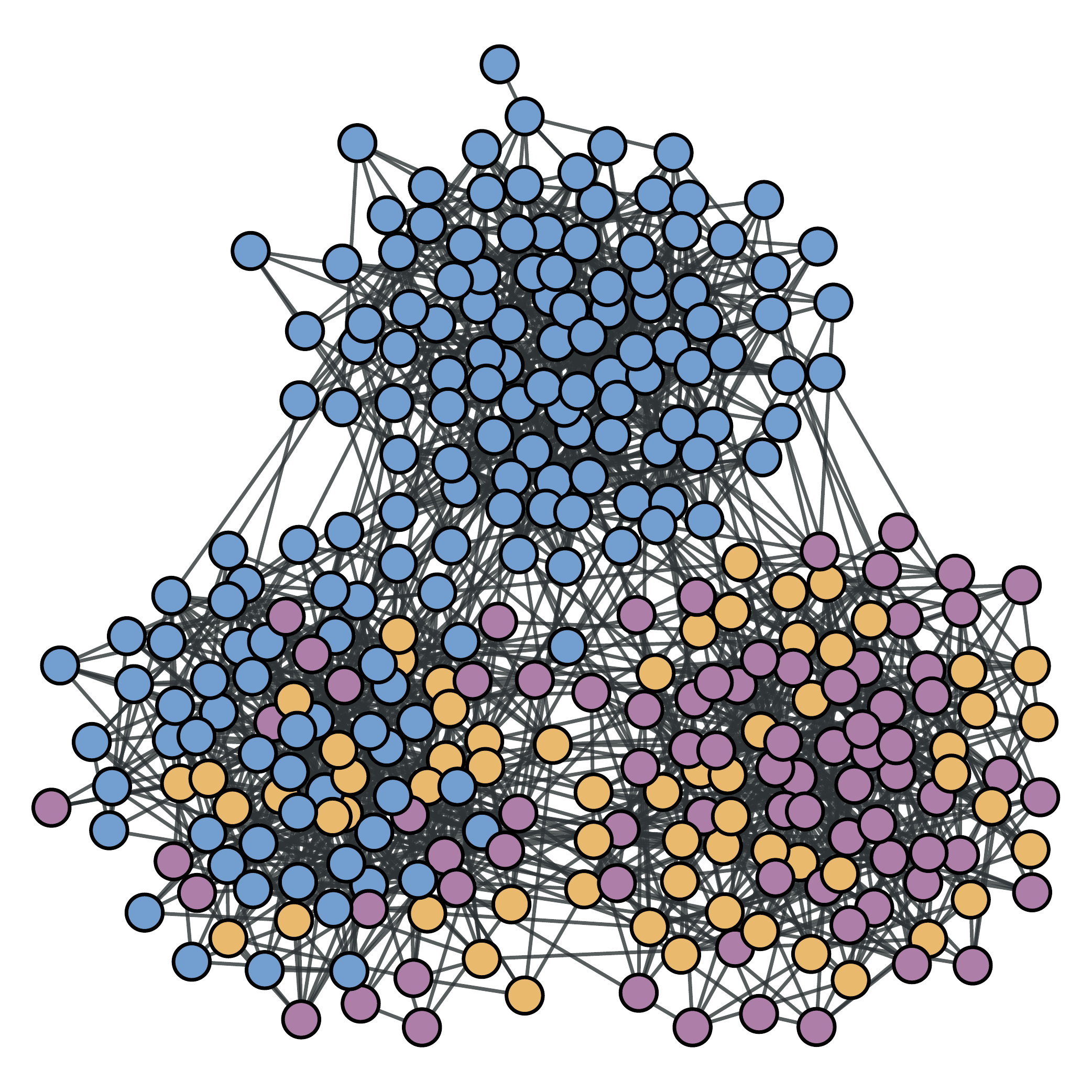}
  \includegraphics[width=.49\columnwidth]{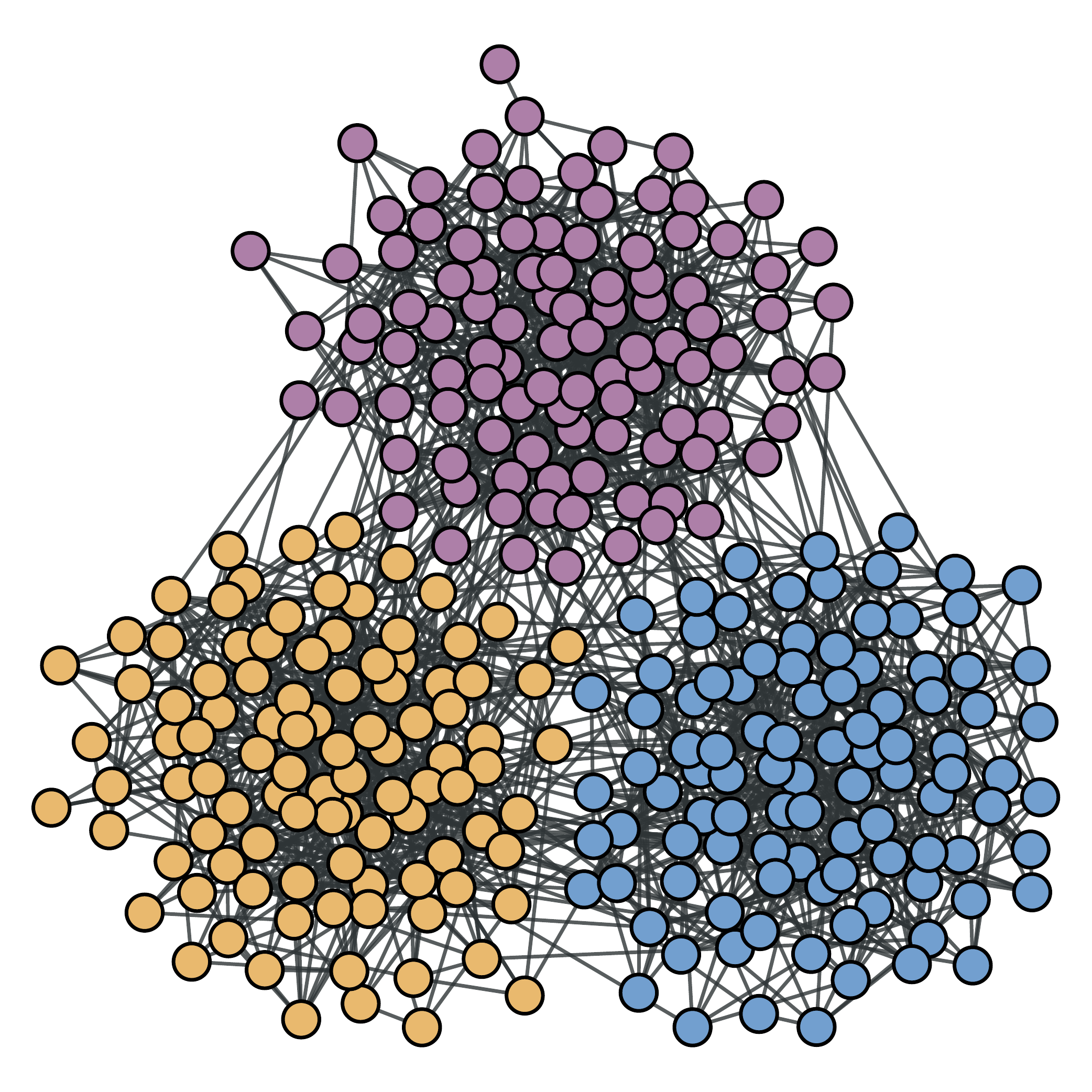} \caption{\label{fig:greedy}\emph{Left:}
  An example of a typical partition obtained by starting with a random
  $B=3$ configuration and applying only greedy moves, until no further
  improvement is possible, for a PP network with $N=300$, $\avg{k}=10$,
  and $c=0.9$.
  \emph{Right:} A typical outcome for the same
  network, with the greedy agglomerative algorithm described in the
  text.}
\end{figure}

Despite its greedy nature, we found that this approach is capable of
almost always avoiding the metastable configurations described
previously, and often comes very close or even exactly to the planted
partition (see Fig.~\ref{fig:greedy}).

The parameters $n_m$, $\sigma$ and $\epsilon$ allow one to choose an
appropriate trade-off between quality and speed. The best results are
obtained for large $n_m$ and small $\sigma$, however these need not to
be chosen fully independently. We found that setting $n_m$ to a
``reasonable'' value such as $10$ or $100$, and selecting $\sigma$ to be
 $2$, $1.1$ or $1.01$ allows one to probe the full quality range
of the algorithm (see below). The choice of the value $\epsilon$ is
interesting, since making $\epsilon=0$ allows one to preserve certain
graph invariants throughout the whole procedure. Since at the first
merging step when $B_i=N$ the $e_{rs}$ matrix is simply the adjacency
matrix, the membership moves with $\epsilon=0$ cannot merge nodes which
belong to different components, or to different partitions in bipartite
networks. It is easy to see that this property is preserved for later
merging steps as well, so they are fully reflected in the final block
structure. We find that very often this is a desired property, and leads
to better block partitions. In situations where it is not desired, it
can be disabled by setting $\epsilon > 0$.

The algorithm above can be turned into a more robust MCMC method by
making $\beta=1$ in the intermediary phase between each merge step, and
waiting sufficiently long for the Markov chain to equilibrate. This is a
slower, but more exact counterpart to the greedy heuristic variant,
which is less susceptible to getting trapped in the metastable states
discussed previously. If one wishes to find the minimum of
$\mathcal{S}_{t/c}$, one can make $\beta\to\infty$ after the chain has
equilibrated, either abruptly (as we do in the results presented in this
paper), or slowly via simulated
annealing~\cite{kirkpatrick_optimization_1983}.

\begin{figure}
  \includegraphics[width=1\columnwidth]{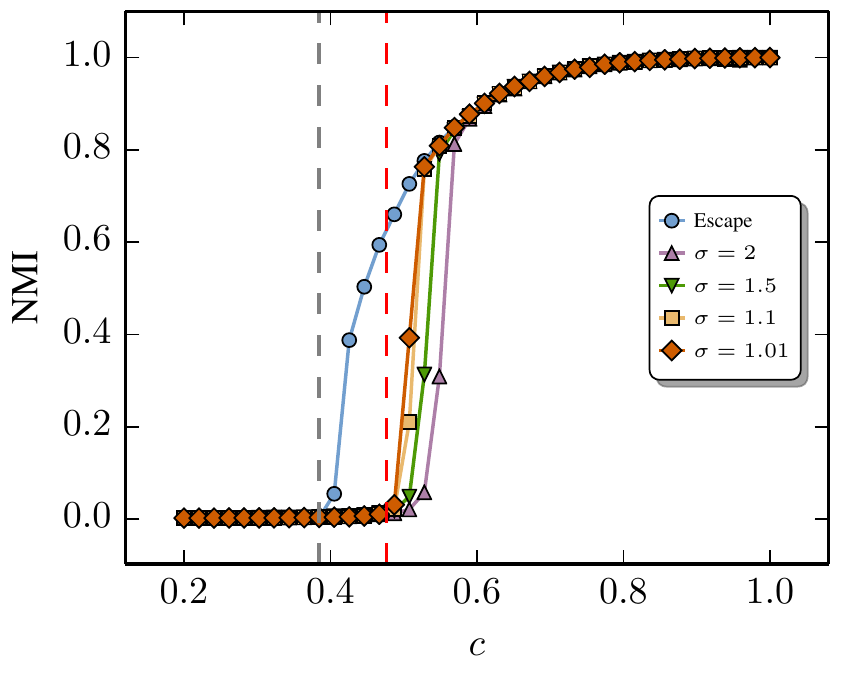}
  \includegraphics[width=1\columnwidth]{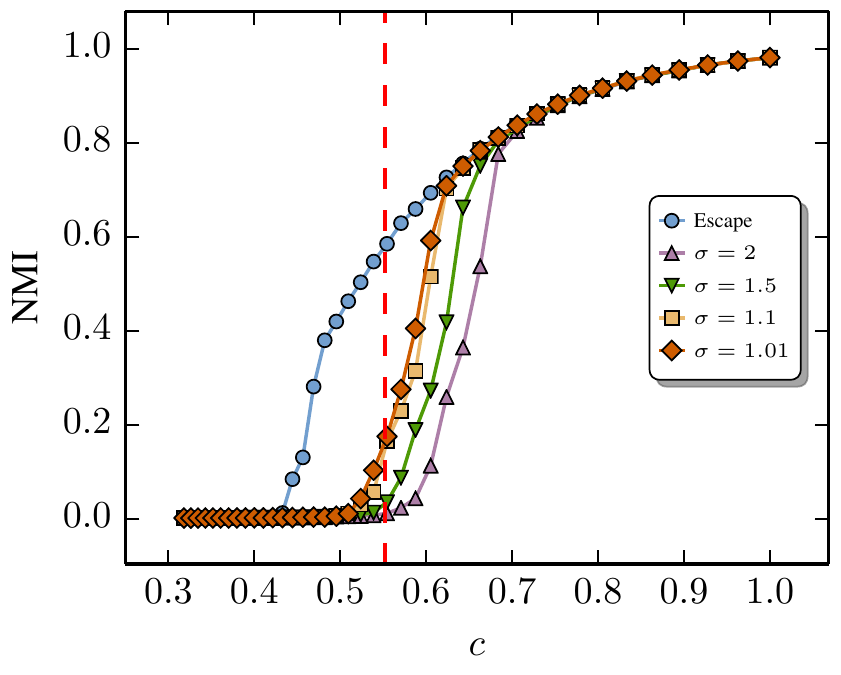}

  \caption{\label{fig:planted}
    Normalized mutual information (NMI) (see footnote~6) between the
    planted and the inferred partitions for (top) the PP model and
    (bottom) the circular multipartite model described in the text, as a
    function of the modular strength $c$, for $N=10^4$ and $B=10$. The
    ``Escape'' curves correspond to MCMC equilibrations starting from
    the planted partition, and the remaining curves to the greedy
    agglomerative heuristic with ratio $\sigma$ shown in the legend, and
    $n_m=10$. All curves are averaged over $20$ independent network
    realizations. The grey vertical dashed line corresponds to the
    detectability threshold $c^*$ for the PP model, and the red dashed
    line to the MDL model selection threshold of Eq.~\ref{eq:mdl}.}
\end{figure}

We can assess the quality of the heuristic method by comparing with
known bounds on the detectability of the PP model. If we have that $N/B
\gg 1$, it can be shown that for $\avg{k} <
[(B-1)/(cB-1)]^2$~\cite{decelle_inference_2011, decelle_asymptotic_2011,
mossel_stochastic_2012}, it is not possible to detect the planted
partition with any method. To emphasize the applicability of the method
for dissortative (or arbitrary) topologies, we also analyze a circular
multipartite block model, with $e_{rs}=2E\left[\left(\delta_{r, s - 1} +
\delta_{r, s + 1}\right)c/2B + (1 - c)/ B^2\right]$, where $c$ controls
the strength of the modular structure, and periodic boundaries are
assumed. In both cases we compare the agglomerative heuristic with MCMC
results starting from the true partition, which represents the best
possible case. As can be seen in Fig.~\ref{fig:planted}, the results
from the optimal MCMC and the heuristic are identical for up to some
values of $c$ which are larger than the actual detectability
threshold. Thus the greedy method falls short of saturating the
detectable parameter region, but behaves badly only for a relatively
small range of $c$, below which it becomes much harder (but not
impossible) to distinguish the observed network from a random graph. To
give a more precise idea of the extent to which the graphs in this
region deviate from a random topology, we compare with a model selection
threshold based on the minimum description length (MDL)
principle~\cite{peixoto_parsimonious_2013},
\begin{equation}\label{eq:mdl}
  \avg{k} > \frac{2\ln B}{I_{t/c}},
\end{equation}
with $I_{t/c} = (S^r_{t/c} - S_{t/c})/E$, where $S_{t/c}^r$ is the
entropy for a fully random graph, with $e_{rs}= 2En_rn_s/N^2$ (or
$e_{rs}= e_re_s/2E$ for the degree-corrected case), and $E \gg B^2$ was
assumed. This criterion is useful when we do not know the correct value
of $B$, and hence cannot rely on minimizing $S_{t/c}$ alone, since it
would always result in a $B=N$ partition. If this condition is not
fulfilled, the inferred partition (even if exact) is discarded in favor
of a fully random graph, since the model parameters in this case cannot
be used to provide a more compact description of the network. From
Fig.~\ref{fig:planted} we see that this threshold lies very close to the
region where the agglomerative algorithm is incapable of discovering the
optimal partition. Hence, in situations where model selection needs to
be performed, any significant improvement to the quality of the
algorithm would be ultimately discarded, at least in these specific
examples. In other situations, where an increased precision close to the
detectability transition is desired, the heuristic should be used only
as a component of the full-fledged MCMC procedure with $\beta=1$, as
described above, which should be able to eventually reach the optimal
configurations, but requires longer running times.

\section{Performance on empirical networks}\label{sec:empirical}

\begin{figure}
  \includegraphics[width=1\columnwidth]{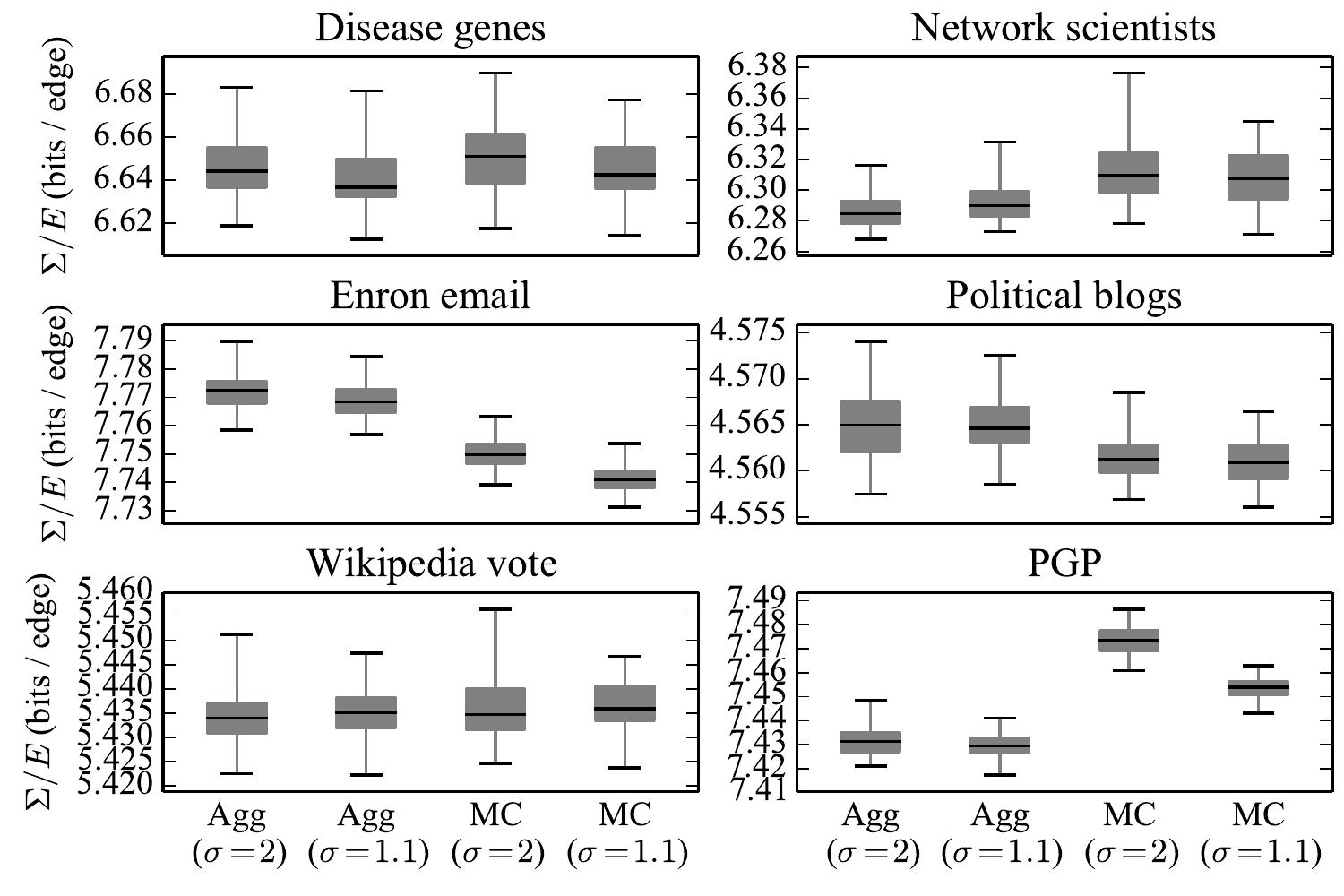}

  \caption{\label{fig:empirical_dl}Description length $\Sigma$ for
  different empirical networks, collected for $100$ independent runs of
  the MCMC algorithm (MC) and the agglomerative heuristic (Agg), for
  different agglomeration ratios $\sigma$. }
\end{figure}

\begin{figure}
  \includegraphics[width=1\columnwidth]{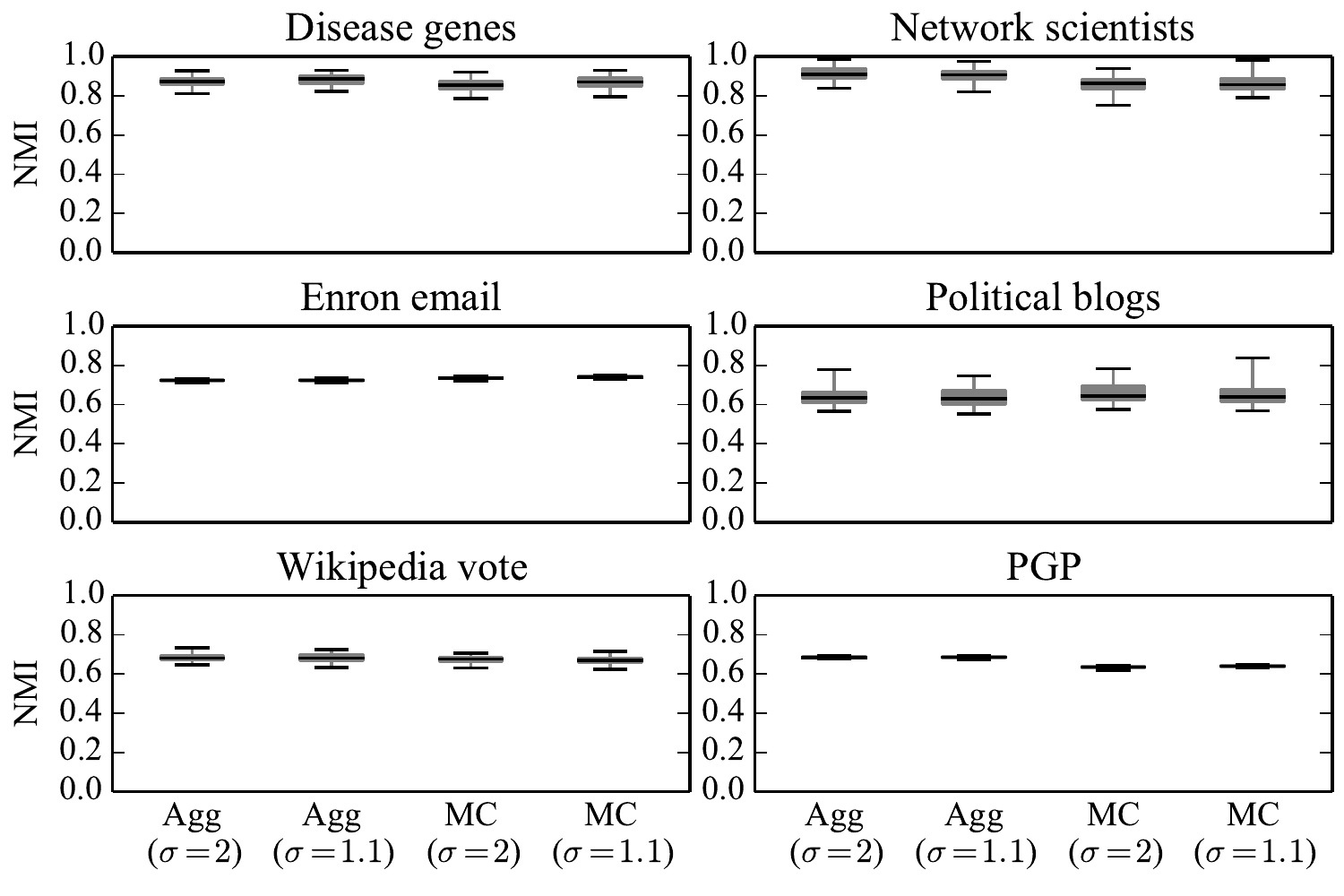}
  \caption{\label{fig:empirical_nmi}
    Normalized mutual information (NMI) between the best overall
    partition and each one collected for $100$ independent runs of the
    MCMC algorithm (MC) and the agglomerative heuristic (Agg), for
    different agglomeration ratios $\sigma$.
}
\end{figure}

\begin{figure}
  \includegraphics[width=.49\columnwidth]{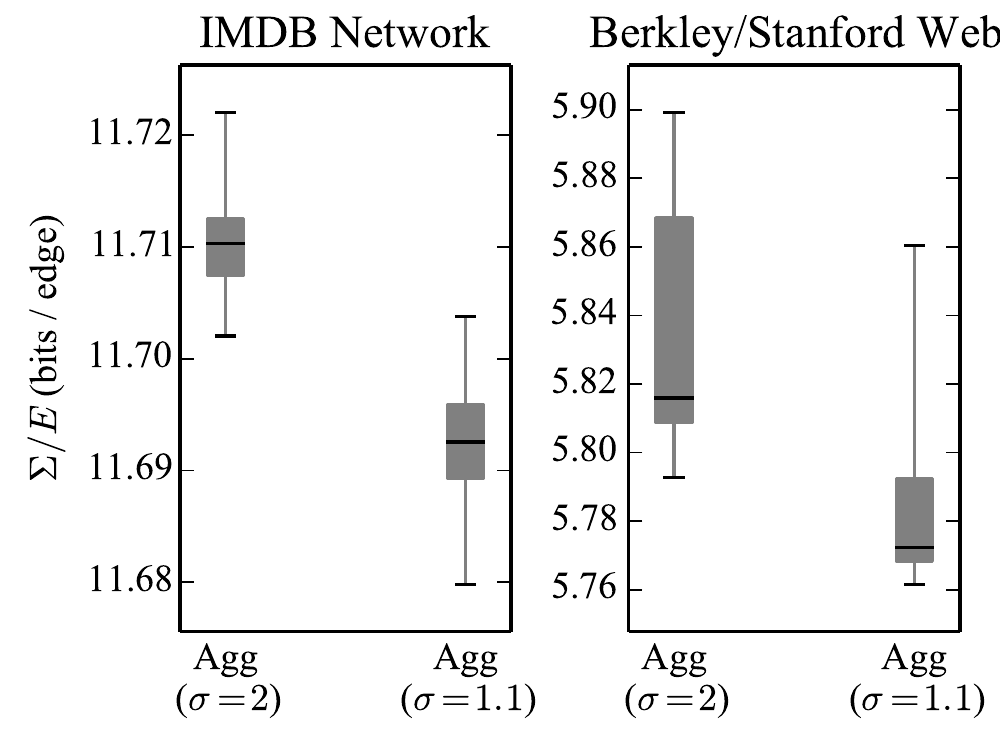}
  \includegraphics[width=.49\columnwidth]{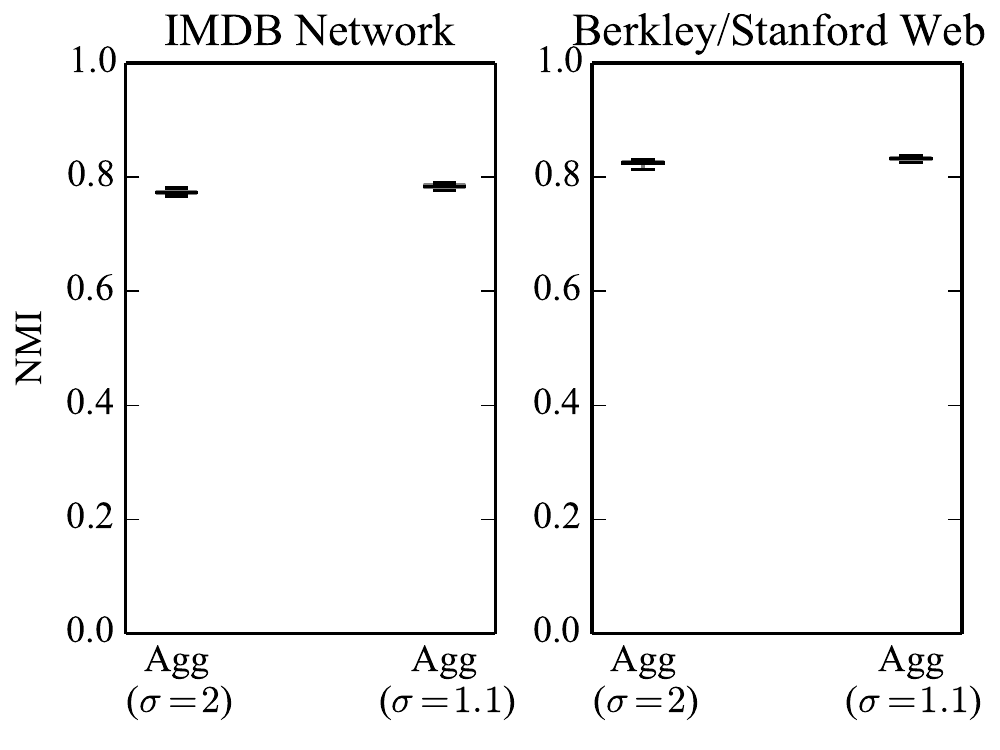}

  \caption{\label{fig:empirical_large}Description length $\Sigma$ for
  different empirical networks, as well the Normalized mutual
  information (NMI) between the best overall partition and each one,
  collected for $100$ independent runs of the agglomerative heuristic,
  for different agglomeration ratios $\sigma$. }
\end{figure}

We have analyzed a few empirical networks to assess the behaivor of the
algorithm in realistic situations.  We have chosen the following
networks: The largest component of coauthorships in network
science~\cite{newman_finding_2006} ($N=379$, $E=914$, undirected), the
human disease gene network~\cite{goh_human_2007} ($N=903$, $E=6,760$,
undirected), the political blog network~\cite{adamic_political_2005}
($N=1,222$, $E=19,021$, directed), the Wikipedia vote
network~\cite{leskovec_signed_2010} ($N=8,298$, $E=103,689$, directed),
the Enron email
network~\cite{klimt_introducing_2004,leskovec_community_2008}
($N=36,692, E=367,662$, undirected), the largest strong component of the
PGP network~\cite{richters_trust_2011} ($N=39,796$, $E=301,498$,
directed), the IMDB film-actor network~\cite{peixoto_parsimonious_2013}
($N=372,547$, $E=1,812,312$, undirected), and the Berkeley/Stanford web
graph~\cite{leskovec_community_2008} ($N=654,782$, $E=7,499,425$,
directed). In all cases we used the degree-corrected model. Since for
these networks the most appropriate value of $B$ is unknown, we
performed model selection using the MDL criterion as described in
Ref.~\cite{peixoto_parsimonious_2013}, where we find the partition which
minimizes the description length $\Sigma = \mathcal{L}_{t/c} +
\mathcal{S}_{t/c}$, where $\mathcal{L}_{t/c}$ is the amount of
information necessary to describe the model parameters, which increases
with $B$~\footnote{As mentioned previously, a more refined MDL method
presented in Ref.~\cite{peixoto_hierarchical_2013} computes
$\mathcal{L}_{t/c}$ via a hierarchical sequence of stochastic block
models, which provides better resolution at the expense of some
additional complexity. But since our objective here is to compare
methods of finding partitions, not model selection, we opt for the
simpler criterion.}. For the networks with moderate size we were capable
of comparing the results with the agglomerative heuristic to those of
the more time consuming MCMC method. Figs.~\ref{fig:empirical_dl}
and~\ref{fig:empirical_large} shown the values of $\Sigma$ after several
runs of each algorithm. It can be observed that the results obtained
with both methods seem largely indistinguishable for some networks
(human diseases, network scientists, and Wikipedia votes), whereas the
MCMC algorithm leads to better results for others (Enron email,
political blogs), and interestingly to worse results for the PGP
network. The better results for MCMC are expected, but the worse result
for the PGP network is not. We can explain this by pointing out that for
that network the average value of $\mathcal{S}_c$ obtained with MCMC for
$\beta=1$ noticeably differs from the minimum possible value. Since we
used an abrupt cooling to $\beta\to\infty$, the MCMC is more likely to
get trapped in a local minimum than the agglomerative heuristic, which
is never allowed to heat up to the $\beta=1$ configurations. MCMC would
probably match, or even improve the heuristic results if,
e.g. simulated annealing would be used to reach the $\beta\to\infty$
region. However, this serves as an example of at least one scenario where
the agglomerative heuristic can lead to even better results, despite
being much faster than MCMC.

Perhaps a more meaningful comparison among the different results is to
determine how the obtained partitions differ from each other. This is
shown in Figs.~\ref{fig:empirical_nmi} and~\ref{fig:empirical_large},
where the normalized mutual information
(NMI)~\footnote{\label{foot:nmi}The NMI is defined as
$2I(\{b_i\},\{c_i\}) / [H(\{b_i\}) + H(\{c_i\})]$, where
$I(\{b_i\},\{c_i\}) = \sum_{rs}p_{bc}(r,s) \ln \left(p_{bc}(r,
s)/p_b(r)p_c(s)\right)$, and $H(\{x_i\}) = -\sum_rp_x(r)\ln p_x(r)$,
where $\{b_i\}$ and $\{c_i\}$ are two partitions of the network.}
between the best partition across all runs of all algorithms and every
other partition found is compared for the two algorithms. Despite
leading to different $\Sigma$ values, the typical partitions found for
each algorithm seem equally far from the (approximated) global maximum,
so the difference in $\Sigma$ can be attributed to minor
differences in the partitions. From this we can conclude the
agglomerative heuristic delivers results comparable to MCMC for many
empirical networks, while being significantly faster.

Note that the NMI values in Fig.~\ref{fig:empirical_nmi} are overall
reasonably high, indicating that the partitions are much more similar
than different, however they are almost never $1$, or very close to it,
except for the smallest networks. This seems to point to a certain
degree of degeneracy of optimal partitions, similar to those reported in
Ref.~\cite{good_performance_2010} for methods based on modularity
maximization. A more detailed analysis of this is needed, but we leave
it to future work.

\section{Conclusion}\label{sec:conclusion}

We have presented an optimized MCMC method~\footnote{An efficient C++
implementation of the algorithm described here is freely available as
part of the graph-tool Python library at
\url{http://graph-tool.skewed.de}.} for inferring stochastic block
models in large networks, which possesses an improved mixing time due to
optimized proposed node membership moves, and an agglomerative procedure
which strongly reduces the likelihood of getting trapped in undesired
metastable states. By increasing the inverse temperature to
$\beta\to\infty$ this method is turned into an agglomerative heuristic,
with a fast algorithmic complexity of $O(N\ln^2 N)$ in sparse
networks. We have shown that although the heuristic does not fully
saturate the detectability range of the MCMC method, it tends to find
indistinguishable partitions for a very large range of parameters of the
generative model, as well as for many empirical networks. The method
also allows for detailed control of the number of blocks $B$ being
inferred, which makes it suitable to be used in conjunction with model
selection techniques~\cite{peixoto_parsimonious_2013,
rosvall_information-theoretic_2007,daudin_mixture_2008,
mariadassou_uncovering_2010, moore_active_2011,
latouche_variational_2012, come_model_2013, peixoto_hierarchical_2013}.

The heuristic method is comparable to the agglomerative algorithm of
Clauset et al~\cite{clauset_finding_2004} (and variants thereof,
e.g. Refs.~\cite{wakita_finding_2007, schuetz_efficient_2008,
schuetz_multistep_2008}), which has the same overall complexity, but is
restricted to finding purely assortative block structures, based on
modularity optimization, and is strictly agglomerative, whereas the
algorithm presented here permits individual node moves between the
blocks at every stage, which allows for the correction of bad merges
done in the earliest stages. It can also be compared to the popular
method of Blondel et al~\cite{blondel_fast_2008}, which is not strictly
agglomerative, but it is also restricted to assortative structures, and
is based on modularity, although it is typically faster than either the
method of Clauset et al and the method presented here.

Both the MCMC method and the greedy heuristic compare favorably to many
statistical inference methods which depend on obtaining the full
marginal probability $\pi_r^i$ that node $i$ belongs to block
$r$~\cite{decelle_inference_2011,decelle_asymptotic_2011,yan_model_2012}. Although
this gives more detailed information on the network structure, it does
so at the expense of much increased algorithmic complexity. For
instance, the belief propagation approach of
Refs.~\cite{decelle_inference_2011,decelle_asymptotic_2011,yan_model_2012},
although it possesses strong optimal properties, requires $O(NB^2)$
operations per update sweep, in addition to an $O(EB)$ memory
complexity. Since in realistic situations the desired value of $B$ is
likely to scale with some power of $N$, this approach quickly becomes
impractical and hinders its application to very large networks, in
contrast to the log-linear complexity in $N$ (independent of $B$) with
the method proposed in this paper.

\bibliographystyle{apsrev4-1} \bibliography{bib}

\begin{thebibliography}{52}%
\makeatletter
\providecommand \@ifxundefined [1]{%
 \@ifx{#1\undefined}
}%
\providecommand \@ifnum [1]{%
 \ifnum #1\expandafter \@firstoftwo
 \else \expandafter \@secondoftwo
 \fi
}%
\providecommand \@ifx [1]{%
 \ifx #1\expandafter \@firstoftwo
 \else \expandafter \@secondoftwo
 \fi
}%
\providecommand \natexlab [1]{#1}%
\providecommand \enquote  [1]{``#1''}%
\providecommand \bibnamefont  [1]{#1}%
\providecommand \bibfnamefont [1]{#1}%
\providecommand \citenamefont [1]{#1}%
\providecommand \href@noop [0]{\@secondoftwo}%
\providecommand \href [0]{\begingroup \@sanitize@url \@href}%
\providecommand \@href[1]{\@@startlink{#1}\@@href}%
\providecommand \@@href[1]{\endgroup#1\@@endlink}%
\providecommand \@sanitize@url [0]{\catcode `\\12\catcode `\$12\catcode
  `\&12\catcode `\#12\catcode `\^12\catcode `\_12\catcode `\%12\relax}%
\providecommand \@@startlink[1]{}%
\providecommand \@@endlink[0]{}%
\providecommand \url  [0]{\begingroup\@sanitize@url \@url }%
\providecommand \@url [1]{\endgroup\@href {#1}{\urlprefix }}%
\providecommand \urlprefix  [0]{URL }%
\providecommand \Eprint [0]{\href }%
\providecommand \doibase [0]{http://dx.doi.org/}%
\providecommand \selectlanguage [0]{\@gobble}%
\providecommand \bibinfo  [0]{\@secondoftwo}%
\providecommand \bibfield  [0]{\@secondoftwo}%
\providecommand \translation [1]{[#1]}%
\providecommand \BibitemOpen [0]{}%
\providecommand \bibitemStop [0]{}%
\providecommand \bibitemNoStop [0]{.\EOS\space}%
\providecommand \EOS [0]{\spacefactor3000\relax}%
\providecommand \BibitemShut  [1]{\csname bibitem#1\endcsname}%
\let\auto@bib@innerbib\@empty
\bibitem [{\citenamefont {Hastings}(2006)}]{hastings_community_2006}%
  \BibitemOpen
  \bibfield  {author} {\bibinfo {author} {\bibfnamefont {M.~B.}\ \bibnamefont
  {Hastings}},\ }\href {\doibase 10.1103/PhysRevE.74.035102} {\bibfield
  {journal} {\bibinfo  {journal} {Physical Review E}\ }\textbf {\bibinfo
  {volume} {74}},\ \bibinfo {pages} {035102} (\bibinfo {year}
  {2006})}\BibitemShut {NoStop}%
\bibitem [{\citenamefont {Garlaschelli}\ and\ \citenamefont
  {Loffredo}(2008)}]{garlaschelli_maximum_2008}%
  \BibitemOpen
  \bibfield  {author} {\bibinfo {author} {\bibfnamefont {D.}~\bibnamefont
  {Garlaschelli}}\ and\ \bibinfo {author} {\bibfnamefont {M.~I.}\ \bibnamefont
  {Loffredo}},\ }\href {\doibase 10.1103/PhysRevE.78.015101} {\bibfield
  {journal} {\bibinfo  {journal} {Physical Review E}\ }\textbf {\bibinfo
  {volume} {78}},\ \bibinfo {pages} {015101} (\bibinfo {year}
  {2008})}\BibitemShut {NoStop}%
\bibitem [{\citenamefont {Newman}\ and\ \citenamefont
  {Leicht}(2007)}]{newman_mixture_2007}%
  \BibitemOpen
  \bibfield  {author} {\bibinfo {author} {\bibfnamefont {M.~E.~J.}\
  \bibnamefont {Newman}}\ and\ \bibinfo {author} {\bibfnamefont {E.~A.}\
  \bibnamefont {Leicht}},\ }\href {\doibase 10.1073/pnas.0610537104} {\bibfield
   {journal} {\bibinfo  {journal} {Proceedings of the National Academy of
  Sciences}\ }\textbf {\bibinfo {volume} {104}},\ \bibinfo {pages} {9564 }
  (\bibinfo {year} {2007})}\BibitemShut {NoStop}%
\bibitem [{\citenamefont {Reichardt}\ and\ \citenamefont
  {White}(2007)}]{reichardt_role_2007}%
  \BibitemOpen
  \bibfield  {author} {\bibinfo {author} {\bibfnamefont {J.}~\bibnamefont
  {Reichardt}}\ and\ \bibinfo {author} {\bibfnamefont {D.~R.}\ \bibnamefont
  {White}},\ }\href {\doibase 10.1140/epjb/e2007-00340-y} {\bibfield  {journal}
  {\bibinfo  {journal} {The European Physical Journal B}\ }\textbf {\bibinfo
  {volume} {60}},\ \bibinfo {pages} {217} (\bibinfo {year} {2007})}\BibitemShut
  {NoStop}%
\bibitem [{\citenamefont {Hofman}\ and\ \citenamefont
  {Wiggins}(2008)}]{hofman_bayesian_2008}%
  \BibitemOpen
  \bibfield  {author} {\bibinfo {author} {\bibfnamefont {J.~M.}\ \bibnamefont
  {Hofman}}\ and\ \bibinfo {author} {\bibfnamefont {C.~H.}\ \bibnamefont
  {Wiggins}},\ }\href {\doibase 10.1103/PhysRevLett.100.258701} {\bibfield
  {journal} {\bibinfo  {journal} {Physical Review Letters}\ }\textbf {\bibinfo
  {volume} {100}},\ \bibinfo {pages} {258701} (\bibinfo {year}
  {2008})}\BibitemShut {NoStop}%
\bibitem [{\citenamefont {Bickel}\ and\ \citenamefont
  {Chen}(2009)}]{bickel_nonparametric_2009}%
  \BibitemOpen
  \bibfield  {author} {\bibinfo {author} {\bibfnamefont {P.~J.}\ \bibnamefont
  {Bickel}}\ and\ \bibinfo {author} {\bibfnamefont {A.}~\bibnamefont {Chen}},\
  }\href {\doibase 10.1073/pnas.0907096106} {\bibfield  {journal} {\bibinfo
  {journal} {Proceedings of the National Academy of Sciences}\ }\textbf
  {\bibinfo {volume} {106}},\ \bibinfo {pages} {21068} (\bibinfo {year}
  {2009})}\BibitemShut {NoStop}%
\bibitem [{\citenamefont {Guimerà}\ and\ \citenamefont
  {Sales-Pardo}(2009)}]{guimera_missing_2009}%
  \BibitemOpen
  \bibfield  {author} {\bibinfo {author} {\bibfnamefont {R.}~\bibnamefont
  {Guimerà}}\ and\ \bibinfo {author} {\bibfnamefont {M.}~\bibnamefont
  {Sales-Pardo}},\ }\href {\doibase 10.1073/pnas.0908366106} {\bibfield
  {journal} {\bibinfo  {journal} {Proceedings of the National Academy of
  Sciences}\ }\textbf {\bibinfo {volume} {106}},\ \bibinfo {pages} {22073 }
  (\bibinfo {year} {2009})}\BibitemShut {NoStop}%
\bibitem [{\citenamefont {Karrer}\ and\ \citenamefont
  {Newman}(2011)}]{karrer_stochastic_2011}%
  \BibitemOpen
  \bibfield  {author} {\bibinfo {author} {\bibfnamefont {B.}~\bibnamefont
  {Karrer}}\ and\ \bibinfo {author} {\bibfnamefont {M.~E.~J.}\ \bibnamefont
  {Newman}},\ }\href {\doibase 10.1103/PhysRevE.83.016107} {\bibfield
  {journal} {\bibinfo  {journal} {Physical Review E}\ }\textbf {\bibinfo
  {volume} {83}},\ \bibinfo {pages} {016107} (\bibinfo {year}
  {2011})}\BibitemShut {NoStop}%
\bibitem [{\citenamefont {Ball}\ \emph {et~al.}(2011)\citenamefont {Ball},
  \citenamefont {Karrer},\ and\ \citenamefont {Newman}}]{ball_efficient_2011}%
  \BibitemOpen
  \bibfield  {author} {\bibinfo {author} {\bibfnamefont {B.}~\bibnamefont
  {Ball}}, \bibinfo {author} {\bibfnamefont {B.}~\bibnamefont {Karrer}}, \ and\
  \bibinfo {author} {\bibfnamefont {M.~E.~J.}\ \bibnamefont {Newman}},\ }\href
  {\doibase 10.1103/PhysRevE.84.036103} {\bibfield  {journal} {\bibinfo
  {journal} {Physical Review E}\ }\textbf {\bibinfo {volume} {84}},\ \bibinfo
  {pages} {036103} (\bibinfo {year} {2011})}\BibitemShut {NoStop}%
\bibitem [{\citenamefont {Reichardt}\ \emph {et~al.}(2011)\citenamefont
  {Reichardt}, \citenamefont {Alamino},\ and\ \citenamefont
  {Saad}}]{reichardt_interplay_2011}%
  \BibitemOpen
  \bibfield  {author} {\bibinfo {author} {\bibfnamefont {J.}~\bibnamefont
  {Reichardt}}, \bibinfo {author} {\bibfnamefont {R.}~\bibnamefont {Alamino}},
  \ and\ \bibinfo {author} {\bibfnamefont {D.}~\bibnamefont {Saad}},\ }\href
  {\doibase 10.1371/journal.pone.0021282} {\bibfield  {journal} {\bibinfo
  {journal} {{PLoS} {ONE}}\ }\textbf {\bibinfo {volume} {6}},\ \bibinfo {pages}
  {e21282} (\bibinfo {year} {2011})}\BibitemShut {NoStop}%
\bibitem [{\citenamefont {Zhu}\ \emph {et~al.}(2012)\citenamefont {Zhu},
  \citenamefont {Yan},\ and\ \citenamefont {Moore}}]{zhu_oriented_2012}%
  \BibitemOpen
  \bibfield  {author} {\bibinfo {author} {\bibfnamefont {Y.}~\bibnamefont
  {Zhu}}, \bibinfo {author} {\bibfnamefont {X.}~\bibnamefont {Yan}}, \ and\
  \bibinfo {author} {\bibfnamefont {C.}~\bibnamefont {Moore}},\ }\href
  {http://arxiv.org/abs/1205.7009} {\bibfield  {journal} {\bibinfo  {journal}
  {{arXiv:1205.7009}}\ } (\bibinfo {year} {2012})}\BibitemShut {NoStop}%
\bibitem [{\citenamefont {Baskerville}\ \emph {et~al.}(2011)\citenamefont
  {Baskerville}, \citenamefont {Dobson}, \citenamefont {Bedford}, \citenamefont
  {Allesina}, \citenamefont {Anderson},\ and\ \citenamefont
  {Pascual}}]{baskerville_spatial_2011}%
  \BibitemOpen
  \bibfield  {author} {\bibinfo {author} {\bibfnamefont {E.~B.}\ \bibnamefont
  {Baskerville}}, \bibinfo {author} {\bibfnamefont {A.~P.}\ \bibnamefont
  {Dobson}}, \bibinfo {author} {\bibfnamefont {T.}~\bibnamefont {Bedford}},
  \bibinfo {author} {\bibfnamefont {S.}~\bibnamefont {Allesina}}, \bibinfo
  {author} {\bibfnamefont {T.~M.}\ \bibnamefont {Anderson}}, \ and\ \bibinfo
  {author} {\bibfnamefont {M.}~\bibnamefont {Pascual}},\ }\href {\doibase
  10.1371/journal.pcbi.1002321} {\bibfield  {journal} {\bibinfo  {journal}
  {{PLoS} Comput Biol}\ }\textbf {\bibinfo {volume} {7}},\ \bibinfo {pages}
  {e1002321} (\bibinfo {year} {2011})}\BibitemShut {NoStop}%
\bibitem [{\citenamefont {Newman}\ and\ \citenamefont
  {Girvan}(2004)}]{newman_finding_2004}%
  \BibitemOpen
  \bibfield  {author} {\bibinfo {author} {\bibfnamefont {M.~E.~J.}\
  \bibnamefont {Newman}}\ and\ \bibinfo {author} {\bibfnamefont
  {M.}~\bibnamefont {Girvan}},\ }\href {\doibase 10.1103/PhysRevE.69.026113}
  {\bibfield  {journal} {\bibinfo  {journal} {Physical Review E}\ }\textbf
  {\bibinfo {volume} {69}},\ \bibinfo {pages} {026113} (\bibinfo {year}
  {2004})}\BibitemShut {NoStop}%
\bibitem [{\citenamefont {Holland}\ \emph {et~al.}(1983)\citenamefont
  {Holland}, \citenamefont {Laskey},\ and\ \citenamefont
  {Leinhardt}}]{holland_stochastic_1983}%
  \BibitemOpen
  \bibfield  {author} {\bibinfo {author} {\bibfnamefont {P.~W.}\ \bibnamefont
  {Holland}}, \bibinfo {author} {\bibfnamefont {K.~B.}\ \bibnamefont {Laskey}},
  \ and\ \bibinfo {author} {\bibfnamefont {S.}~\bibnamefont {Leinhardt}},\
  }\href {\doibase 16/0378-8733(83)90021-7} {\bibfield  {journal} {\bibinfo
  {journal} {Social Networks}\ }\textbf {\bibinfo {volume} {5}},\ \bibinfo
  {pages} {109} (\bibinfo {year} {1983})}\BibitemShut {NoStop}%
\bibitem [{\citenamefont {Fienberg}\ \emph {et~al.}(1985)\citenamefont
  {Fienberg}, \citenamefont {Meyer},\ and\ \citenamefont
  {Wasserman}}]{fienberg_statistical_1985}%
  \BibitemOpen
  \bibfield  {author} {\bibinfo {author} {\bibfnamefont {S.~E.}\ \bibnamefont
  {Fienberg}}, \bibinfo {author} {\bibfnamefont {M.~M.}\ \bibnamefont {Meyer}},
  \ and\ \bibinfo {author} {\bibfnamefont {S.~S.}\ \bibnamefont {Wasserman}},\
  }\href {\doibase 10.2307/2288040} {\bibfield  {journal} {\bibinfo  {journal}
  {Journal of the American Statistical Association}\ }\textbf {\bibinfo
  {volume} {80}},\ \bibinfo {pages} {51} (\bibinfo {year} {1985})}\BibitemShut
  {NoStop}%
\bibitem [{\citenamefont {Faust}\ and\ \citenamefont
  {Wasserman}(1992)}]{faust_blockmodels:_1992}%
  \BibitemOpen
  \bibfield  {author} {\bibinfo {author} {\bibfnamefont {K.}~\bibnamefont
  {Faust}}\ and\ \bibinfo {author} {\bibfnamefont {S.}~\bibnamefont
  {Wasserman}},\ }\href {\doibase 16/0378-8733(92)90013-W} {\bibfield
  {journal} {\bibinfo  {journal} {Social Networks}\ }\textbf {\bibinfo {volume}
  {14}},\ \bibinfo {pages} {5} (\bibinfo {year} {1992})}\BibitemShut {NoStop}%
\bibitem [{\citenamefont {Anderson}\ \emph {et~al.}(1992)\citenamefont
  {Anderson}, \citenamefont {Wasserman},\ and\ \citenamefont
  {Faust}}]{anderson_building_1992}%
  \BibitemOpen
  \bibfield  {author} {\bibinfo {author} {\bibfnamefont {C.~J.}\ \bibnamefont
  {Anderson}}, \bibinfo {author} {\bibfnamefont {S.}~\bibnamefont {Wasserman}},
  \ and\ \bibinfo {author} {\bibfnamefont {K.}~\bibnamefont {Faust}},\ }\href
  {\doibase 16/0378-8733(92)90017-2} {\bibfield  {journal} {\bibinfo  {journal}
  {Social Networks}\ }\textbf {\bibinfo {volume} {14}},\ \bibinfo {pages} {137}
  (\bibinfo {year} {1992})}\BibitemShut {NoStop}%
\bibitem [{\citenamefont {Fortunato}(2010)}]{fortunato_community_2010}%
  \BibitemOpen
  \bibfield  {author} {\bibinfo {author} {\bibfnamefont {S.}~\bibnamefont
  {Fortunato}},\ }\href {\doibase 16/j.physrep.2009.11.002} {\bibfield
  {journal} {\bibinfo  {journal} {Physics Reports}\ }\textbf {\bibinfo {volume}
  {486}},\ \bibinfo {pages} {75} (\bibinfo {year} {2010})}\BibitemShut
  {NoStop}%
\bibitem [{\citenamefont
  {Peixoto}(2013{\natexlab{a}})}]{peixoto_parsimonious_2013}%
  \BibitemOpen
  \bibfield  {author} {\bibinfo {author} {\bibfnamefont {T.~P.}\ \bibnamefont
  {Peixoto}},\ }\href {\doibase 10.1103/PhysRevLett.110.148701} {\bibfield
  {journal} {\bibinfo  {journal} {Physical Review Letters}\ }\textbf {\bibinfo
  {volume} {110}},\ \bibinfo {pages} {148701} (\bibinfo {year}
  {2013}{\natexlab{a}})}\BibitemShut {NoStop}%
\bibitem [{\citenamefont {Rosvall}\ and\ \citenamefont
  {Bergstrom}(2007)}]{rosvall_information-theoretic_2007}%
  \BibitemOpen
  \bibfield  {author} {\bibinfo {author} {\bibfnamefont {M.}~\bibnamefont
  {Rosvall}}\ and\ \bibinfo {author} {\bibfnamefont {C.~T.}\ \bibnamefont
  {Bergstrom}},\ }\href {\doibase 10.1073/pnas.0611034104} {\bibfield
  {journal} {\bibinfo  {journal} {Proceedings of the National Academy of
  Sciences}\ }\textbf {\bibinfo {volume} {104}},\ \bibinfo {pages} {7327}
  (\bibinfo {year} {2007})}\BibitemShut {NoStop}%
\bibitem [{\citenamefont {Daudin}\ \emph {et~al.}(2008)\citenamefont {Daudin},
  \citenamefont {Picard},\ and\ \citenamefont {Robin}}]{daudin_mixture_2008}%
  \BibitemOpen
  \bibfield  {author} {\bibinfo {author} {\bibfnamefont {J.-J.}\ \bibnamefont
  {Daudin}}, \bibinfo {author} {\bibfnamefont {F.}~\bibnamefont {Picard}}, \
  and\ \bibinfo {author} {\bibfnamefont {S.}~\bibnamefont {Robin}},\ }\href
  {\doibase 10.1007/s11222-007-9046-7} {\bibfield  {journal} {\bibinfo
  {journal} {Statistics and Computing}\ }\textbf {\bibinfo {volume} {18}},\
  \bibinfo {pages} {173} (\bibinfo {year} {2008})}\BibitemShut {NoStop}%
\bibitem [{\citenamefont {Mariadassou}\ \emph {et~al.}(2010)\citenamefont
  {Mariadassou}, \citenamefont {Robin},\ and\ \citenamefont
  {Vacher}}]{mariadassou_uncovering_2010}%
  \BibitemOpen
  \bibfield  {author} {\bibinfo {author} {\bibfnamefont {M.}~\bibnamefont
  {Mariadassou}}, \bibinfo {author} {\bibfnamefont {S.}~\bibnamefont {Robin}},
  \ and\ \bibinfo {author} {\bibfnamefont {C.}~\bibnamefont {Vacher}},\ }\href
  {\doibase 10.1214/10-AOAS361} {\bibfield  {journal} {\bibinfo  {journal} {The
  Annals of Applied Statistics}\ }\textbf {\bibinfo {volume} {4}},\ \bibinfo
  {pages} {715} (\bibinfo {year} {2010})},\ \bibinfo {note} {mathematical
  Reviews number ({MathSciNet):} {MR2758646}}\BibitemShut {NoStop}%
\bibitem [{\citenamefont {Moore}\ \emph {et~al.}(2011)\citenamefont {Moore},
  \citenamefont {Yan}, \citenamefont {Zhu}, \citenamefont {Rouquier},\ and\
  \citenamefont {Lane}}]{moore_active_2011}%
  \BibitemOpen
  \bibfield  {author} {\bibinfo {author} {\bibfnamefont {C.}~\bibnamefont
  {Moore}}, \bibinfo {author} {\bibfnamefont {X.}~\bibnamefont {Yan}}, \bibinfo
  {author} {\bibfnamefont {Y.}~\bibnamefont {Zhu}}, \bibinfo {author}
  {\bibfnamefont {J.-B.}\ \bibnamefont {Rouquier}}, \ and\ \bibinfo {author}
  {\bibfnamefont {T.}~\bibnamefont {Lane}},\ }in\ \href {\doibase
  10.1145/2020408.2020552} {\emph {\bibinfo {booktitle} {Proceedings of the
  17th {ACM} {SIGKDD} international conference on Knowledge discovery and data
  mining}}},\ \bibinfo {series and number} {{KDD} '11}\ (\bibinfo  {publisher}
  {{ACM}},\ \bibinfo {address} {New York, {NY}, {USA}},\ \bibinfo {year}
  {2011})\ p.\ \bibinfo {pages} {841–849}\BibitemShut {NoStop}%
\bibitem [{\citenamefont {Latouche}\ \emph {et~al.}(2012)\citenamefont
  {Latouche}, \citenamefont {Birmele},\ and\ \citenamefont
  {Ambroise}}]{latouche_variational_2012}%
  \BibitemOpen
  \bibfield  {author} {\bibinfo {author} {\bibfnamefont {P.}~\bibnamefont
  {Latouche}}, \bibinfo {author} {\bibfnamefont {E.}~\bibnamefont {Birmele}}, \
  and\ \bibinfo {author} {\bibfnamefont {C.}~\bibnamefont {Ambroise}},\ }\href
  {http://smj.sagepub.com/content/12/1/93.short} {\bibfield  {journal}
  {\bibinfo  {journal} {Statistical Modelling}\ }\textbf {\bibinfo {volume}
  {12}},\ \bibinfo {pages} {93–115} (\bibinfo {year} {2012})}\BibitemShut
  {NoStop}%
\bibitem [{\citenamefont {Côme}\ and\ \citenamefont
  {Latouche}(2013)}]{come_model_2013}%
  \BibitemOpen
  \bibfield  {author} {\bibinfo {author} {\bibfnamefont {E.}~\bibnamefont
  {Côme}}\ and\ \bibinfo {author} {\bibfnamefont {P.}~\bibnamefont
  {Latouche}},\ }\href {http://arxiv.org/abs/1303.2962} {\emph {\bibinfo
  {title} {Model selection and clustering in stochastic block models with the
  exact integrated complete data likelihood}}},\ \bibinfo {type} {{arXiv}
  e-print}\ \bibinfo {number} {1303.2962}\ (\bibinfo {year} {2013})\BibitemShut
  {NoStop}%
\bibitem [{\citenamefont
  {Peixoto}(2013{\natexlab{b}})}]{peixoto_hierarchical_2013}%
  \BibitemOpen
  \bibfield  {author} {\bibinfo {author} {\bibfnamefont {T.~P.}\ \bibnamefont
  {Peixoto}},\ }\href {http://arxiv.org/abs/1310.4377} {\emph {\bibinfo {title}
  {Hierarchical block structures and high-resolution model selection in large
  networks}}},\ \bibinfo {type} {{arXiv} e-print}\ \bibinfo {number}
  {1310.4377}\ (\bibinfo {year} {2013})\BibitemShut {NoStop}%
\bibitem [{\citenamefont {Decelle}\ \emph
  {et~al.}(2011{\natexlab{a}})\citenamefont {Decelle}, \citenamefont
  {Krzakala}, \citenamefont {Moore},\ and\ \citenamefont
  {Zdeborová}}]{decelle_inference_2011}%
  \BibitemOpen
  \bibfield  {author} {\bibinfo {author} {\bibfnamefont {A.}~\bibnamefont
  {Decelle}}, \bibinfo {author} {\bibfnamefont {F.}~\bibnamefont {Krzakala}},
  \bibinfo {author} {\bibfnamefont {C.}~\bibnamefont {Moore}}, \ and\ \bibinfo
  {author} {\bibfnamefont {L.}~\bibnamefont {Zdeborová}},\ }\href {\doibase
  10.1103/PhysRevLett.107.065701} {\bibfield  {journal} {\bibinfo  {journal}
  {Physical Review Letters}\ }\textbf {\bibinfo {volume} {107}},\ \bibinfo
  {pages} {065701} (\bibinfo {year} {2011}{\natexlab{a}})}\BibitemShut
  {NoStop}%
\bibitem [{\citenamefont {Decelle}\ \emph
  {et~al.}(2011{\natexlab{b}})\citenamefont {Decelle}, \citenamefont
  {Krzakala}, \citenamefont {Moore},\ and\ \citenamefont
  {Zdeborová}}]{decelle_asymptotic_2011}%
  \BibitemOpen
  \bibfield  {author} {\bibinfo {author} {\bibfnamefont {A.}~\bibnamefont
  {Decelle}}, \bibinfo {author} {\bibfnamefont {F.}~\bibnamefont {Krzakala}},
  \bibinfo {author} {\bibfnamefont {C.}~\bibnamefont {Moore}}, \ and\ \bibinfo
  {author} {\bibfnamefont {L.}~\bibnamefont {Zdeborová}},\ }\href {\doibase
  10.1103/PhysRevE.84.066106} {\bibfield  {journal} {\bibinfo  {journal}
  {Physical Review E}\ }\textbf {\bibinfo {volume} {84}},\ \bibinfo {pages}
  {066106} (\bibinfo {year} {2011}{\natexlab{b}})}\BibitemShut {NoStop}%
\bibitem [{\citenamefont {Mossel}\ \emph {et~al.}(2012)\citenamefont {Mossel},
  \citenamefont {Neeman},\ and\ \citenamefont {Sly}}]{mossel_stochastic_2012}%
  \BibitemOpen
  \bibfield  {author} {\bibinfo {author} {\bibfnamefont {E.}~\bibnamefont
  {Mossel}}, \bibinfo {author} {\bibfnamefont {J.}~\bibnamefont {Neeman}}, \
  and\ \bibinfo {author} {\bibfnamefont {A.}~\bibnamefont {Sly}},\ }\href
  {http://arxiv.org/abs/1202.1499} {\bibfield  {journal} {\bibinfo  {journal}
  {{arXiv:1202.1499}}\ } (\bibinfo {year} {2012})}\BibitemShut {NoStop}%
\bibitem [{\citenamefont {Reichardt}\ and\ \citenamefont
  {Leone}(2008)}]{reichardt_detectable_2008}%
  \BibitemOpen
  \bibfield  {author} {\bibinfo {author} {\bibfnamefont {J.}~\bibnamefont
  {Reichardt}}\ and\ \bibinfo {author} {\bibfnamefont {M.}~\bibnamefont
  {Leone}},\ }\href {\doibase 10.1103/PhysRevLett.101.078701} {\bibfield
  {journal} {\bibinfo  {journal} {Physical Review Letters}\ }\textbf {\bibinfo
  {volume} {101}},\ \bibinfo {pages} {078701} (\bibinfo {year}
  {2008})}\BibitemShut {NoStop}%
\bibitem [{\citenamefont {Hu}\ \emph {et~al.}(2012)\citenamefont {Hu},
  \citenamefont {Ronhovde},\ and\ \citenamefont {Nussinov}}]{hu_phase_2012}%
  \BibitemOpen
  \bibfield  {author} {\bibinfo {author} {\bibfnamefont {D.}~\bibnamefont
  {Hu}}, \bibinfo {author} {\bibfnamefont {P.}~\bibnamefont {Ronhovde}}, \ and\
  \bibinfo {author} {\bibfnamefont {Z.}~\bibnamefont {Nussinov}},\ }\href
  {\doibase 10.1080/14786435.2011.616547} {\bibfield  {journal} {\bibinfo
  {journal} {Philosophical Magazine}\ }\textbf {\bibinfo {volume} {92}},\
  \bibinfo {pages} {406} (\bibinfo {year} {2012})}\BibitemShut {NoStop}%
\bibitem [{\citenamefont {Clauset}\ \emph {et~al.}(2004)\citenamefont
  {Clauset}, \citenamefont {Newman},\ and\ \citenamefont
  {Moore}}]{clauset_finding_2004}%
  \BibitemOpen
  \bibfield  {author} {\bibinfo {author} {\bibfnamefont {A.}~\bibnamefont
  {Clauset}}, \bibinfo {author} {\bibfnamefont {M.~E.~J.}\ \bibnamefont
  {Newman}}, \ and\ \bibinfo {author} {\bibfnamefont {C.}~\bibnamefont
  {Moore}},\ }\href {\doibase 10.1103/PhysRevE.70.066111} {\bibfield  {journal}
  {\bibinfo  {journal} {Physical Review E}\ }\textbf {\bibinfo {volume} {70}},\
  \bibinfo {pages} {066111} (\bibinfo {year} {2004})}\BibitemShut {NoStop}%
\bibitem [{\citenamefont {Blondel}\ \emph {et~al.}(2008)\citenamefont
  {Blondel}, \citenamefont {Guillaume}, \citenamefont {Lambiotte},\ and\
  \citenamefont {Lefebvre}}]{blondel_fast_2008}%
  \BibitemOpen
  \bibfield  {author} {\bibinfo {author} {\bibfnamefont {V.~D.}\ \bibnamefont
  {Blondel}}, \bibinfo {author} {\bibfnamefont {J.-L.}\ \bibnamefont
  {Guillaume}}, \bibinfo {author} {\bibfnamefont {R.}~\bibnamefont
  {Lambiotte}}, \ and\ \bibinfo {author} {\bibfnamefont {E.}~\bibnamefont
  {Lefebvre}},\ }\href {\doibase 10.1088/1742-5468/2008/10/P10008} {\bibfield
  {journal} {\bibinfo  {journal} {Journal of Statistical Mechanics: Theory and
  Experiment}\ }\textbf {\bibinfo {volume} {2008}},\ \bibinfo {pages} {P10008}
  (\bibinfo {year} {2008})}\BibitemShut {NoStop}%
\bibitem [{\citenamefont {Bianconi}(2009)}]{bianconi_entropy_2009}%
  \BibitemOpen
  \bibfield  {author} {\bibinfo {author} {\bibfnamefont {G.}~\bibnamefont
  {Bianconi}},\ }\href {\doibase 10.1103/PhysRevE.79.036114} {\bibfield
  {journal} {\bibinfo  {journal} {Physical Review E}\ }\textbf {\bibinfo
  {volume} {79}},\ \bibinfo {pages} {036114} (\bibinfo {year}
  {2009})}\BibitemShut {NoStop}%
\bibitem [{\citenamefont {Peixoto}(2012)}]{peixoto_entropy_2012}%
  \BibitemOpen
  \bibfield  {author} {\bibinfo {author} {\bibfnamefont {T.~P.}\ \bibnamefont
  {Peixoto}},\ }\href {\doibase 10.1103/PhysRevE.85.056122} {\bibfield
  {journal} {\bibinfo  {journal} {Physical Review E}\ }\textbf {\bibinfo
  {volume} {85}},\ \bibinfo {pages} {056122} (\bibinfo {year}
  {2012})}\BibitemShut {NoStop}%
\bibitem [{\citenamefont {Fortunato}\ and\ \citenamefont
  {Barthélemy}(2007)}]{fortunato_resolution_2007}%
  \BibitemOpen
  \bibfield  {author} {\bibinfo {author} {\bibfnamefont {S.}~\bibnamefont
  {Fortunato}}\ and\ \bibinfo {author} {\bibfnamefont {M.}~\bibnamefont
  {Barthélemy}},\ }\href {\doibase 10.1073/pnas.0605965104} {\bibfield
  {journal} {\bibinfo  {journal} {Proceedings of the National Academy of
  Sciences}\ }\textbf {\bibinfo {volume} {104}},\ \bibinfo {pages} {36}
  (\bibinfo {year} {2007})}\BibitemShut {NoStop}%
\bibitem [{\citenamefont {Metropolis}\ \emph {et~al.}(1953)\citenamefont
  {Metropolis}, \citenamefont {Rosenbluth}, \citenamefont {Rosenbluth},
  \citenamefont {Teller},\ and\ \citenamefont
  {Teller}}]{metropolis_equation_1953}%
  \BibitemOpen
  \bibfield  {author} {\bibinfo {author} {\bibfnamefont {N.}~\bibnamefont
  {Metropolis}}, \bibinfo {author} {\bibfnamefont {A.~W.}\ \bibnamefont
  {Rosenbluth}}, \bibinfo {author} {\bibfnamefont {M.~N.}\ \bibnamefont
  {Rosenbluth}}, \bibinfo {author} {\bibfnamefont {A.~H.}\ \bibnamefont
  {Teller}}, \ and\ \bibinfo {author} {\bibfnamefont {E.}~\bibnamefont
  {Teller}},\ }\href {\doibase 10.1063/1.1699114} {\bibfield  {journal}
  {\bibinfo  {journal} {The Journal of Chemical Physics}\ }\textbf {\bibinfo
  {volume} {21}},\ \bibinfo {pages} {1087} (\bibinfo {year}
  {1953})}\BibitemShut {NoStop}%
\bibitem [{\citenamefont {Hastings}(1970)}]{hastings_monte_1970}%
  \BibitemOpen
  \bibfield  {author} {\bibinfo {author} {\bibfnamefont {W.~K.}\ \bibnamefont
  {Hastings}},\ }\href {\doibase 10.1093/biomet/57.1.97} {\bibfield  {journal}
  {\bibinfo  {journal} {Biometrika}\ }\textbf {\bibinfo {volume} {57}},\
  \bibinfo {pages} {97 } (\bibinfo {year} {1970})}\BibitemShut {NoStop}%
\bibitem [{\citenamefont {Condon}\ and\ \citenamefont
  {Karp}(2001)}]{condon_algorithms_2001}%
  \BibitemOpen
  \bibfield  {author} {\bibinfo {author} {\bibfnamefont {A.}~\bibnamefont
  {Condon}}\ and\ \bibinfo {author} {\bibfnamefont {R.~M.}\ \bibnamefont
  {Karp}},\ }\href {\doibase
  10.1002/1098-2418(200103)18:2<116::AID-RSA1001>3.0.CO;2-2} {\bibfield
  {journal} {\bibinfo  {journal} {Random Structures \& Algorithms}\ }\textbf
  {\bibinfo {volume} {18}},\ \bibinfo {pages} {116–140} (\bibinfo {year}
  {2001})}\BibitemShut {NoStop}%
\bibitem [{\citenamefont {Kirkpatrick}\ \emph {et~al.}(1983)\citenamefont
  {Kirkpatrick}, \citenamefont {Gelatt~Jr},\ and\ \citenamefont
  {Vecchi}}]{kirkpatrick_optimization_1983}%
  \BibitemOpen
  \bibfield  {author} {\bibinfo {author} {\bibfnamefont {S.}~\bibnamefont
  {Kirkpatrick}}, \bibinfo {author} {\bibfnamefont {C.~D.}\ \bibnamefont
  {Gelatt~Jr}}, \ and\ \bibinfo {author} {\bibfnamefont {M.~P.}\ \bibnamefont
  {Vecchi}},\ }\href@noop {} {\bibfield  {journal} {\bibinfo  {journal}
  {Science}\ }\textbf {\bibinfo {volume} {220}},\ \bibinfo {pages} {671}
  (\bibinfo {year} {1983})}\BibitemShut {NoStop}%
\bibitem [{\citenamefont {Newman}(2006)}]{newman_finding_2006}%
  \BibitemOpen
  \bibfield  {author} {\bibinfo {author} {\bibfnamefont {M.~E.~J.}\
  \bibnamefont {Newman}},\ }\href {\doibase 10.1103/PhysRevE.74.036104}
  {\bibfield  {journal} {\bibinfo  {journal} {Physical Review E}\ }\textbf
  {\bibinfo {volume} {74}},\ \bibinfo {pages} {036104} (\bibinfo {year}
  {2006})}\BibitemShut {NoStop}%
\bibitem [{\citenamefont {Goh}\ \emph {et~al.}(2007)\citenamefont {Goh},
  \citenamefont {Cusick}, \citenamefont {Valle}, \citenamefont {Childs},
  \citenamefont {Vidal},\ and\ \citenamefont {Barabási}}]{goh_human_2007}%
  \BibitemOpen
  \bibfield  {author} {\bibinfo {author} {\bibfnamefont {K.~I.}\ \bibnamefont
  {Goh}}, \bibinfo {author} {\bibfnamefont {M.~E.}\ \bibnamefont {Cusick}},
  \bibinfo {author} {\bibfnamefont {D.}~\bibnamefont {Valle}}, \bibinfo
  {author} {\bibfnamefont {B.}~\bibnamefont {Childs}}, \bibinfo {author}
  {\bibfnamefont {M.}~\bibnamefont {Vidal}}, \ and\ \bibinfo {author}
  {\bibfnamefont {A.~L.}\ \bibnamefont {Barabási}},\ }\href
  {http://www.pnas.org/content/104/21/8685.short} {\bibfield  {journal}
  {\bibinfo  {journal} {Proceedings of the National Academy of Sciences}\
  }\textbf {\bibinfo {volume} {104}},\ \bibinfo {pages} {8685} (\bibinfo {year}
  {2007})}\BibitemShut {NoStop}%
\bibitem [{\citenamefont {Adamic}\ and\ \citenamefont
  {Glance}(2005)}]{adamic_political_2005}%
  \BibitemOpen
  \bibfield  {author} {\bibinfo {author} {\bibfnamefont {L.~A.}\ \bibnamefont
  {Adamic}}\ and\ \bibinfo {author} {\bibfnamefont {N.}~\bibnamefont
  {Glance}},\ }in\ \href {\doibase 10.1145/1134271.1134277} {\emph {\bibinfo
  {booktitle} {Proceedings of the 3rd international workshop on Link
  discovery}}},\ \bibinfo {series and number} {{LinkKDD} '05}\ (\bibinfo
  {publisher} {{ACM}},\ \bibinfo {address} {New York, {NY}, {USA}},\ \bibinfo
  {year} {2005})\ p.\ \bibinfo {pages} {36–43}\BibitemShut {NoStop}%
\bibitem [{\citenamefont {Leskovec}\ \emph {et~al.}(2010)\citenamefont
  {Leskovec}, \citenamefont {Huttenlocher},\ and\ \citenamefont
  {Kleinberg}}]{leskovec_signed_2010}%
  \BibitemOpen
  \bibfield  {author} {\bibinfo {author} {\bibfnamefont {J.}~\bibnamefont
  {Leskovec}}, \bibinfo {author} {\bibfnamefont {D.}~\bibnamefont
  {Huttenlocher}}, \ and\ \bibinfo {author} {\bibfnamefont {J.}~\bibnamefont
  {Kleinberg}},\ }in\ \href {\doibase 10.1145/1753326.1753532} {\emph {\bibinfo
  {booktitle} {Proceedings of the {SIGCHI} Conference on Human Factors in
  Computing Systems}}},\ \bibinfo {series and number} {{CHI} '10}\ (\bibinfo
  {publisher} {{ACM}},\ \bibinfo {address} {New York, {NY}, {USA}},\ \bibinfo
  {year} {2010})\ p.\ \bibinfo {pages} {1361–1370}\BibitemShut {NoStop}%
\bibitem [{\citenamefont {Klimt}\ and\ \citenamefont
  {Yang}(2004)}]{klimt_introducing_2004}%
  \BibitemOpen
  \bibfield  {author} {\bibinfo {author} {\bibfnamefont {B.}~\bibnamefont
  {Klimt}}\ and\ \bibinfo {author} {\bibfnamefont {Y.}~\bibnamefont {Yang}},\
  }in\ \href@noop {} {\emph {\bibinfo {booktitle} {{CEAS}}}}\ (\bibinfo {year}
  {2004})\BibitemShut {NoStop}%
\bibitem [{\citenamefont {Leskovec}\ \emph {et~al.}(2008)\citenamefont
  {Leskovec}, \citenamefont {Lang}, \citenamefont {Dasgupta},\ and\
  \citenamefont {Mahoney}}]{leskovec_community_2008}%
  \BibitemOpen
  \bibfield  {author} {\bibinfo {author} {\bibfnamefont {J.}~\bibnamefont
  {Leskovec}}, \bibinfo {author} {\bibfnamefont {K.~J.}\ \bibnamefont {Lang}},
  \bibinfo {author} {\bibfnamefont {A.}~\bibnamefont {Dasgupta}}, \ and\
  \bibinfo {author} {\bibfnamefont {M.~W.}\ \bibnamefont {Mahoney}},\ }\href
  {http://arxiv.org/abs/0810.1355} {\bibfield  {journal} {\bibinfo  {journal}
  {{arXiv:0810.1355}}\ } (\bibinfo {year} {2008})}\BibitemShut {NoStop}%
\bibitem [{\citenamefont {Richters}\ and\ \citenamefont
  {Peixoto}(2011)}]{richters_trust_2011}%
  \BibitemOpen
  \bibfield  {author} {\bibinfo {author} {\bibfnamefont {O.}~\bibnamefont
  {Richters}}\ and\ \bibinfo {author} {\bibfnamefont {T.~P.}\ \bibnamefont
  {Peixoto}},\ }\href {\doibase 10.1371/journal.pone.0018384} {\bibfield
  {journal} {\bibinfo  {journal} {{PLoS} {ONE}}\ }\textbf {\bibinfo {volume}
  {6}},\ \bibinfo {pages} {e18384} (\bibinfo {year} {2011})}\BibitemShut
  {NoStop}%
\bibitem [{\citenamefont {Good}\ \emph {et~al.}(2010)\citenamefont {Good},
  \citenamefont {de~Montjoye},\ and\ \citenamefont
  {Clauset}}]{good_performance_2010}%
  \BibitemOpen
  \bibfield  {author} {\bibinfo {author} {\bibfnamefont {B.~H.}\ \bibnamefont
  {Good}}, \bibinfo {author} {\bibfnamefont {Y.-A.}\ \bibnamefont
  {de~Montjoye}}, \ and\ \bibinfo {author} {\bibfnamefont {A.}~\bibnamefont
  {Clauset}},\ }\href {\doibase 10.1103/PhysRevE.81.046106} {\bibfield
  {journal} {\bibinfo  {journal} {Physical Review E}\ }\textbf {\bibinfo
  {volume} {81}},\ \bibinfo {pages} {046106} (\bibinfo {year}
  {2010})}\BibitemShut {NoStop}%
\bibitem [{\citenamefont {Wakita}\ and\ \citenamefont
  {Tsurumi}(2007)}]{wakita_finding_2007}%
  \BibitemOpen
  \bibfield  {author} {\bibinfo {author} {\bibfnamefont {K.}~\bibnamefont
  {Wakita}}\ and\ \bibinfo {author} {\bibfnamefont {T.}~\bibnamefont
  {Tsurumi}},\ }in\ \href {\doibase 10.1145/1242572.1242805} {\emph {\bibinfo
  {booktitle} {Proceedings of the 16th International Conference on World Wide
  Web}}},\ \bibinfo {series and number} {{WWW} '07}\ (\bibinfo  {publisher}
  {{ACM}},\ \bibinfo {address} {New York, {NY}, {USA}},\ \bibinfo {year}
  {2007})\ p.\ \bibinfo {pages} {1275–1276}\BibitemShut {NoStop}%
\bibitem [{\citenamefont {Schuetz}\ and\ \citenamefont
  {Caflisch}(2008{\natexlab{a}})}]{schuetz_efficient_2008}%
  \BibitemOpen
  \bibfield  {author} {\bibinfo {author} {\bibfnamefont {P.}~\bibnamefont
  {Schuetz}}\ and\ \bibinfo {author} {\bibfnamefont {A.}~\bibnamefont
  {Caflisch}},\ }\href {\doibase 10.1103/PhysRevE.77.046112} {\bibfield
  {journal} {\bibinfo  {journal} {Physical Review E}\ }\textbf {\bibinfo
  {volume} {77}},\ \bibinfo {pages} {046112} (\bibinfo {year}
  {2008}{\natexlab{a}})}\BibitemShut {NoStop}%
\bibitem [{\citenamefont {Schuetz}\ and\ \citenamefont
  {Caflisch}(2008{\natexlab{b}})}]{schuetz_multistep_2008}%
  \BibitemOpen
  \bibfield  {author} {\bibinfo {author} {\bibfnamefont {P.}~\bibnamefont
  {Schuetz}}\ and\ \bibinfo {author} {\bibfnamefont {A.}~\bibnamefont
  {Caflisch}},\ }\href {\doibase 10.1103/PhysRevE.78.026112} {\bibfield
  {journal} {\bibinfo  {journal} {Physical Review E}\ }\textbf {\bibinfo
  {volume} {78}},\ \bibinfo {pages} {026112} (\bibinfo {year}
  {2008}{\natexlab{b}})}\BibitemShut {NoStop}%
\bibitem [{\citenamefont {Yan}\ \emph {et~al.}(2012)\citenamefont {Yan},
  \citenamefont {Jensen}, \citenamefont {Krzakala}, \citenamefont {Moore},
  \citenamefont {Shalizi}, \citenamefont {Zdeborova}, \citenamefont {Zhang},\
  and\ \citenamefont {Zhu}}]{yan_model_2012}%
  \BibitemOpen
  \bibfield  {author} {\bibinfo {author} {\bibfnamefont {X.}~\bibnamefont
  {Yan}}, \bibinfo {author} {\bibfnamefont {J.~E.}\ \bibnamefont {Jensen}},
  \bibinfo {author} {\bibfnamefont {F.}~\bibnamefont {Krzakala}}, \bibinfo
  {author} {\bibfnamefont {C.}~\bibnamefont {Moore}}, \bibinfo {author}
  {\bibfnamefont {C.~R.}\ \bibnamefont {Shalizi}}, \bibinfo {author}
  {\bibfnamefont {L.}~\bibnamefont {Zdeborova}}, \bibinfo {author}
  {\bibfnamefont {P.}~\bibnamefont {Zhang}}, \ and\ \bibinfo {author}
  {\bibfnamefont {Y.}~\bibnamefont {Zhu}},\ }\href
  {http://arxiv.org/abs/1207.3994} {\bibfield  {journal} {\bibinfo  {journal}
  {{arXiv:1207.3994}}\ } (\bibinfo {year} {2012})}\BibitemShut {NoStop}%
\end{thebibliography}%

\end{document}